\documentclass[12pt]{JHEP3}

\usepackage{epsfig}
\usepackage{subfigure}
\usepackage{wrapfig}

\usepackage{amsmath}
\usepackage{amssymb}
\usepackage{cite}
\usepackage{array}

\newcommand{\arctanh}{{\rm arctanh}\,}
\newcommand{\arcsinh}{{\rm arcsinh}\,}

\newcommand{\beqs}{\begin{equation*}}
\newcommand{\beq}{\begin{equation}}

\newcommand{\eeqs}{\end{equation*}}
\newcommand{\eeq}{\end{equation}}

\newcommand{\beqas}{\begin{eqnarray*}}
\newcommand{\beqa}{\begin{eqnarray}}

\newcommand{\eeqas}{\end{eqnarray*}}
\newcommand{\eeqa}{\end{eqnarray}}

\newcommand{\eq}[2]{\begin{equation} #1 \label{#2} \end{equation}}

\newcommand{\eps}{\varepsilon}
\newcommand{\al}{\alpha}

\newcommand{\ga}{\gamma}
\newcommand{\de}{\delta}
\newcommand{\om}{\omega}

\newcommand{\la}{\lambda}

\newcommand{\De}{\Delta}
\newcommand{\Om}{\Omega}

\newcommand{\blist}{\begin{itemize}}

\newcommand{\elist}{\end{itemize}}

\providecommand{\href}[2]{#2}

\DeclareFontFamily{OT1}{rsfs}{}
\DeclareFontShape{OT1}{rsfs}{m}{n}{ <-7> rsfs5 <7-10> rsfs7 <10->rsfs10}{} 
\DeclareMathAlphabet{\mycal}{OT1}{rsfs}{m}{n}

\newcommand{\diff}{\extd}

\DeclareMathOperator{\extdm}{d}
\newcommand{\extd}{\extdm \!}

\newcommand{\tac}{{\mathcal T}}
\newcommand{\scalarf}{\tac}

\title{An action for the exact string black hole}

\author{D.\ Grumiller\\ Institute for Theoretical Physics, University of Leipzig\\ Augustusplatz 10-11, D-04109 Leipzig, Germany\\ E-mail: \email{grumiller@itp.uni-leipzig.de}}

\abstract{
A local action is constructed describing the exact string black hole discovered by
Dijkgraaf, Verlinde and Verlinde in 1992. It turns out to be a special 2D Maxwell-dilaton gravity theory, linear in curvature and field strength. 
Two constants of motion exist: mass $M\geq 1$, determined by the level $k$, and $U(1)$-charge $Q\geq 0$, determined by the value of the dilaton at the origin. 
ADM mass, Hawking temperature $T_H\propto\sqrt{1-1/M}$
 and Bekenstein--Hawking entropy are derived and studied in detail. 
Winding/momentum mode duality implies the existence of a similar action, arising from a branch ambiguity, which describes the exact string naked singularity. In the strong coupling limit the solution dual to $AdS_2$ is found to be the 5D Schwarzschild black hole. Some applications to black hole thermodynamics and 2D string theory are discussed and generalizations -- supersymmetric extension, coupling to matter and critical collapse, quantization -- are pointed out.}


\preprint{LU-ITP-2005/001}

\dedicated{Dedicated to Wolfgang Kummer on occasion of his Emeritierung}

\begin{document}

\section{Introduction}

 The exact string black hole (ESBH) was discovered by Dijkgraaf, Verlinde and Verlinde (DVV) more than a decade ago \cite{Dijkgraaf:1992ba}. The construction of an action for it which does not display non-localities or higher order derivatives 
is a challenging open problem in the context of 2D string theory. 
The purpose of this paper is to solve it.

There are several advantages of having such an action available: the main point of the ESBH is its non-perturbative aspect, i.e., it is believed to be valid to all orders in the string-coupling $\alpha^\prime$. Thus, a corresponding action will capture non-perturbative features of string theory and will allow, among other things, for the first time a thorough discussion of ADM mass, Hawking temperature and Bekenstein--Hawking entropy of the ESBH which otherwise requires some ad-hoc assumption. Moreover, once an action is at our disposal an exact path integral quantization may be performed. 
A more detailed exposition of these issues and other applications will be postponed until the conclusions.

At the perturbative level actions approximating the ESBH are known: to lowest order in $\alpha^\prime$ an action emerges the classical solutions of which describe the Witten BH \cite{Witten:1991yr,Mandal:1991tz,Elitzur:1991cb}, which in turn inspired the CGHS model \cite{Callan:1992rs}, a 2D dilaton gravity model with scalar matter that has been used as a toy model for BH evaporation. Pushing perturbative considerations further Tseytlin was able to show that up to 3 loops the ESBH is consistent with sigma model conformal invariance \cite{Tseytlin:1991ht}; in the supersymmetric case this holds even at 4 loops \cite{Jack:1993mk}. In the strong coupling regime the ESBH asymptotes to the Jackiw-Teitelboim (JT) model \cite{Teitelboim:1983ux}. 
The exact conformal field theory (CFT) methods used in \cite{Dijkgraaf:1992ba}, based upon the ${\rm \it SL}(2,\mathbb{R})/U(1)$ gauged Wess-Zumino-Witten model, imply the dependence of the ESBH solutions on the level $k$. A different (somewhat more direct) derivation leading to the same results for dilaton and metric was presented in \cite{Tseytlin:1992ri} (see also \cite{Bars:1993zf}). For a comprehensive history and more references ref.~\cite{Becker:1994vd} may be consulted.

In the notation of \cite{Kazakov:2001pj} for Euclidean signature the line element of the ESBH 
discovered by DVV is given by
\eq{
\extd s^2=f^2(x)\extd\tau^2+\extd x^2\,,
}{eq:dvv1}
with
\eq{
f(x)=\frac{\tanh{(bx)}}{\sqrt{1-p\tanh^2{(bx)}}}\,.
}{eq:dvv2}
Physical scales are adjusted by the parameter $b\in\mathbb{R}^+$ which has dimension of
inverse length. The corresponding expression for the dilaton,  
\eq{
\phi=\phi_0-\ln{\cosh{(bx)}}-\frac{1}{4}\ln{(1-p\tanh^2{(bx)})}\,,
}{eq:dvv3}
contains an integration constant $\phi_0$.
Additionally, there are the following relations between constants, 
string-coupling $\al^\prime$, level $k$ and dimension $D$ of string target 
space:
\eq{
\alpha^\prime b^2=\frac{1}{k-2}\,, \hspace{0.5cm}
p:=\frac{2}{k}=\frac{2\alpha^\prime b^2}{1+2\alpha^\prime b^2}\,, \hspace{0.5cm} 
D-26+6\alpha^\prime b^2=0\,.
}{eq:dvv4}
For $D=2$ one obtains $p=8/9$, but like in the original work 
\cite{Dijkgraaf:1992ba} we will treat general values of $p\in(0;1)$ and 
consider the limits $p\to 0$ and $p\to 1$ separately: for $p=0$ one recovers
the Witten BH; for $p=1$ the JT model 
is obtained. Both limits exhibit singular features: 
for all $p\in(0;1)$ the solution is regular 
globally, asymptotically flat and exactly one Killing horizon exists. However, 
for $p=0$ a curvature singularity (screened by a horizon) appears and for $p=1$ 
space-time fails to be asymptotically flat.
In the present work exclusively the Minkowskian version of (\ref{eq:dvv1})
\eq{
\extd s^2=f^2(x)\extd\tau^2-\extd x^2\,,
}{eq:dvv5}
will be needed.
The maximally extended space-time of this geometry has been studied by
Perry and Teo \cite{Perry:1993ry} and by Yi \cite{Yi:1993gh}.

Winding/momentum mode duality implies the existence of a dual solution, the Exact String Naked Singularity (ESNS), which 
can be acquired most easily by replacing $bx\to bx+i\pi/2$, entailing in all 
formulas above the substitutions 
\eq{
\sinh\to i\cosh\,,\hspace{0.5cm}\cosh\to i\sinh\,.
}{dvv:dual}

This concludes the brief recollection of the main properties of the ESBH/ESNS relevant for the present work. The task is now clear, if slightly ambitious: we are seeking an action the classical solutions of which yield \eqref{eq:dvv2}-\eqref{eq:dvv5}. Some prejudices concerning the action may be helpful: It has to be a 2D action. It may depend on the scale parameter $b$, but not on the constant $\phi_0$ which should emerge as a constant of motion. It should functionally depend on the metric, the dilaton and eventual auxiliary fields. It should be diffeomorphism invariant and local Lorentz invariant. The absence of non-localities and non-polynomial derivative interactions is crucial. It would be splendid if no propagating physical degrees of freedom (PPDOF) were present and marvellous if the action described not only the ESBH but also, by some ``simple'' duality transformation, the ESNS. Last but not least one has to circumvent the no-go result of ref.~\cite{Grumiller:2002md} by relaxing at least one of its premises.

This paper is organized as follows: for sake of self-containment section \ref{se:2} recalls some of the main results of Maxwell-dilaton gravity in 2D in the first order formalism: various formulations of the action (subsection \ref{se:2.1}), coupling to abelian gauge fields (subsection \ref{se:2.2}), and how to obtain all classical solutions (subsection \ref{se:2.3}). This section may be skipped by readers familiar with that formalism. Section \ref{se:3} contains the main part of the paper: the presentation of the action (subsection \ref{se:3.1}), the proof of equivalence of line element and dilaton to the ESBH (subsection \ref{se:3.2}), and a discussion of the action as well as different representations thereof (subsection \ref{se:3.3}). Equipped with such an action thermodynamical properties may be discussed with ease (section \ref{se:4}): (ADM) mass (subsection \ref{se:4.1}), (Hawking) temperature (subsection \ref{se:4.2}), and (Bekenstein--Hawking) entropy (\ref{se:4.3}). The extensive conclusions in section \ref{se:6} reveal physical features, applications and generalizations, 
and compare with the literature. 
The appendices are devoted to historical remarks. They are recommended to readers interested in a bottom-up construction of the action who may wish to consult them before reading the main statement in section \ref{se:3}: the no-go result is recapitulated in appendix \ref{app:B.1} and the crucial idea of introducing an abelian Maxwell field is put into historical context in appendix \ref{app:B.2}.

\section{Recapitulation of 2D dilaton gravity}\label{se:2}

The purpose of this brief summary of well-known results is to provide a self-contained introduction to dilaton gravity in the first order formalism and to fix the notation.\footnote{The sign of the curvature scalar $r$ has been fixed conveniently such that $r>0$ for $dS_2$. This is the only difference to the notations used in ref.~\cite{Grumiller:2002nm}.} 
For background information and additional references the extensive review \cite{Grumiller:2002nm} may be consulted. Supplementary material providing relations to non-linear algebras may be found in appendix \ref{app:B.2}.

\subsection{Geometry and actions in 2D}\label{se:2.1}

For various reasons, some of which will become apparent while obtaining all classical solutions, it is very convenient to employ the first order formalism: $e^a=e^a_\mu dx^\mu$ is the
dyad 1-form dual to $E_a$ -- i.e.\ $e^a(E_b)=\de^a_b$. Latin indices refer to an anholonomic frame, Greek indices to a holonomic one. The 1-form
$\omega$ represents the  spin-connection $\om^a{}_b=\eps^a{}_b\om$
with  the totally antisymmetric Levi-Civit{\'a} symbol $\eps_{ab}$ ($\eps_{01}=+1$). With the
flat metric $\eta_{ab}$ in light-cone coordinates
($\eta_{+-}=1=\eta_{-+}$, $\eta_{++}=0=\eta_{--}$) it reads $\eps^\pm{}_\pm=\pm 1$. The torsion 2-form is given by 
$T^\pm=(\extd\pm\omega)\wedge e^\pm$. The curvature 2-form $R^a{}_b$ can be represented by the 2-form $R$ defined by 
$R^a{}_b=\eps^a{}_b R$, $R=\extd\om$. The volume 2-form is denoted by $\epsilon = e^+\wedge e^-$. Signs and factors of the Hodge-$\ast$ operation are defined by $\ast\epsilon=1$. 
Since the Einstein-Hilbert action $\int_{\mathcal{M}_2}  R\propto(1-\gamma)$ yields just the Euler number for a surface with genus $\gamma$ one has to generalize it appropriately to generate equations of motion (EOM). The simplest idea is to introduce a Lagrange multiplier for curvature, $X$, also known as ``dilaton field'', and an arbitrary potential thereof, $V(X)$, in the action $\int_{\mathcal{M}_2}  \left(XR+\epsilon V(X)\right)$. 
Having introduced curvature it is natural to consider torsion as well. By analogy the first order gravity action \cite{Schaller:1994es}
\eq{
S^{\rm (1)}= \int_{\mathcal{M}_2}  \left[X_aT^a+XR+\epsilon\mathcal{V} (X^aX_a,X)\right]
}{eq:FOG}
can be motivated where $X_a$ are the Lagrange multipliers for torsion. It encompasses essentially all known dilaton theories in 2D. Actually, for most practical purposes the potential takes the simpler form
\begin{equation}
  \label{eq:pot}
  \mathcal{V} (X^aX_a,X) = X^+X^- U(X) + V(X)\,.
\end{equation}
The action \eqref{eq:FOG} is equivalent to the frequently used second order action \cite{Russo:1992yg,Banks:1991mk,Odintsov:1991qu}
\begin{equation}
\label{eq:GDT}
S^{(2)}=\int_{\mathcal{M}_2} \extd^{2}x\, \sqrt{-g}\; \left[ X \frac{-r}{2}-\frac{U(X)}{2}\; (\nabla X)^{2}+V(X)\; \right] \, ,
\end{equation}
with the same functions $U,V$ as in \eqref{eq:pot}. The curvature scalar $r$ and covariant derivative $\nabla$ are associated with the Levi-Civit\'a connection related to the metric $g_{\mu\nu}$, the determinant of which is denoted by $g$. If $\om$ is torsion-free $r\propto\ast R$.  In the absence of matter there are no PPDOF.

There is another intriguing re-interpretation of \eqref{eq:FOG}: defining $X^I=(X,X^+,X^-)$ and $A_I=(\om,e^-,e^+)$ the Poisson-sigma model (PSM) action emerges \cite{Schaller:1994es}
\begin{equation}
  \label{eq:PSMaction}
S^{(\rm PSM)}=\int_{\mathcal{M}_2} \left[\extd X^I\wedge A_I + \frac12 P^{IJ} A_J\wedge A_I\right]\,,
\end{equation}
provided the Poisson tensor is chosen as
\begin{equation}
  \label{eq:poissontensor}
  P^{IJ}=\left(\begin{array}{ccc}
 0    & X^+          & -X^- \\
 -X^+ & 0            & \mathcal{V} \\
 X^-  & -\mathcal{V} & 0
\end{array}\right)\,.
\end{equation}
Being a Poisson tensor it is not only anti-symmetric but it also fulfills the Jacobi identity
\begin{equation}
  \label{eq:jacobi}
    P^{IL}\partial _{L}P^{JK}+ \mbox{\it perm}\left( IJK\right) = 0\,.
\end{equation}
Such a reformulation is advantageous because e.g.~the existence of a Casimir function may be deduced immediately from \eqref{eq:poissontensor}. It turns out that this Casimir function is related to a conserved quantity, ``the mass'', which has been found in previous second order studies of dilaton gravity \cite{Banks:1991mk,Frolov:1992xx,Mann:1993yv} as well as in the first order formulation  \cite{Grosse:1992vc}. The PSM perspective on 2D dilaton gravity is summarized in \cite{Strobl:1999wv}.

Finally, it should be mentioned that in the second order formalism often the dilaton field $\phi$, with 
\begin{equation}
  \label{eq:dilatondilaton}
  X=e^{-2\phi}\,,
\end{equation}
is employed. This brings \eqref{eq:GDT} into the well-known form
\begin{equation}
  \label{eq:GDTotherdilaton}
  S^{(2')}=-\frac12\int_{\mathcal{M}_2} \extd^{2}x\, \sqrt{-g}\; e^{-2\phi} \left[ r + \hat{U}(\phi)\; (\nabla\phi)^{2} \; +\hat{V}(\phi) \right] \, ,
\end{equation}
where the new potentials $\hat{U}$, $\hat{V}$ are related to the old ones via
\begin{equation}
  \label{eq:newpotentials}
  \hat{U}=4 e^{-2\phi}U(e^{-2\phi})\,,\quad\hat{V}=- 2e^{2\phi}V(e^{-2\phi})\,.
\end{equation}
Two prominent examples are the Witten BH with
\begin{equation}
  \label{eq:potentialwittenbh}
  U(X)=-\frac{1}{X}\,,\quad V(X)=-2b^2X\,,\quad\rightarrow\quad\hat{U}(\phi)=-4\,,\quad\hat{V}=+4b^2\,,
\end{equation}
and the JT model\footnote{Here it is presented only for negative cosmological constant, i.e., $AdS_2$. The $dS_2$ case may be obtained by changing the sign in the definition of $V$ in \eqref{eq:potentialsJT}.} with
\begin{equation}
  \label{eq:potentialsJT}
  U(X)=0\,,\quad V(X)= -b^2 X\,,\quad\rightarrow\quad\hat{U}(\phi)=0\,,\quad\hat{V}=+2b^2\,.
\end{equation}
The scale parameter $b\in\mathbb{R}^+$ defines the physical units and is essentially irrelevant. 

\subsection{Coupling to an abelian gauge field}\label{se:2.2}

It is straightforward to generalize \eqref{eq:FOG} to a Maxwell-dilaton first order action,
\begin{equation}
  \label{eq:fogu1}
  S^{\rm (MD1)}= \int_{\mathcal{M}_2}  \left[X_aT^a+XR+BF+\epsilon\mathcal{V} (X^aX_a,X,B)\right]\,,
\end{equation}
where $B$ is an additional scalar field and $F=\extd A$ is the field strength 2-form, being the differential of the gauge field 1-form $A$. Variation with respect to $A$ immediately establishes a constant of motion,\footnote{In the PSM language adding an abelian gauge field means adding another row and column of zeros to the Poisson tensor \eqref{eq:poissontensor}. Thus, its rank is unchanged and the dimension of its kernel is increased by 1. Therefore, an additional Casimir function exists: the constant of motion in \eqref{eq:casimiru1}.}
\begin{equation}
  \label{eq:casimiru1}
  \extd B=0\,,\quad\rightarrow\quad B = Q\,,
\end{equation}
where $Q$ is some real constant, the $U(1)$ charge. Variation with respect to $B$ may produce a relation that allows to express $B$ as a function of the dilaton and the dual field strength $\ast F$. This need not be the case, however. 

The result \eqref{eq:casimiru1} implies that the solution of the remaining EOM reduces to the case without Maxwell field. One just has to replace $B$ by its on-shell value $Q$ in the potential $\mathcal{V}$. Before discussing how to solve these remaining equations in the next subsection, two examples will be provided.

\paragraph{Example 1: Standard Maxwell-dilaton models} Suppose that $\mathcal{V}=X^+X^-U(X)+V(X)+\frac12 f(X)B^2$. Then, variation with respect to $B$ gives
\begin{equation}
  \label{eq:eomB}
  B=-\frac{\ast F}{f(X)}\,.
\end{equation}
Inserting this back into the action yields the standard geometric part plus the following term
\begin{equation}
  \label{eq:Ladd}
  S^{(\rm add)}=-\frac{1}{2}\int_{\mathcal{M}_2}\frac{\ast F F}{f(X)}\,,
\end{equation}
which is an ordinary Maxwell term with nonminimal coupling to the dilaton via $f(X)$. Alternatively, one can use the on-shell condition $B=Q$, thus obtaining a purely geometric action with an effective potential $\mathcal{V}=X^+X^-U(X)+V(X)+\frac{Q^2}{2} f(X)$. A typical example is spherically reduced gravity, i.e., the Reissner-Nordstr\"om BH, where $f(X)\propto 1/X$.

\paragraph{Example 2: Specific non-standard models} For $\mathcal{V}=V(X)+f(X)B$ variation with respect to $B$ yields
\begin{equation}
  \label{eq:eomBCS}
  f(X)=-\ast F\,.  
\end{equation}
Provided $f(X)$ is invertible the dilaton $X$ may be expressed on-shell as a function of the dual field strength. For the Kaluza-Klein reduced gravitational Chern-Simons theory \cite{Guralnik:2003we} this observation turned out to be pivotal for a successful application of first order gravity methods \cite{Grumiller:2003ad} and supersymmetrization \cite{Bergamin:2004me}.

\subsection{All classical solutions}\label{se:2.3}

It is useful to introduce the following combinations of the potentials $U$ and $V$:
\begin{equation}
  \label{eq:wI}
  I(X):=\exp{\int^X U(y)\extd y}\,,\quad \tilde{w}(X):=\int^XI(y)V(y)\extd y
\end{equation}
The integration constants may be absorbed, respectively, by rescalings and shifts of the ``mass'', see equation \eqref{eq:c} below. Under dilaton dependent conformal transformations $X^a\to X^a/\Om$, $e^a\to e^a\Om$, $\om\to\om+X_ae^a\extd\,\ln{\Om}/\extd X$ equation \eqref{eq:FOG} is mapped to a new action of the same type with transformed potentials $\tilde{U}$, $\tilde{V}$. Hence, it is not invariant. It turns out that only the combination $\tilde{w}(X)$ as defined in \eqref{eq:wI} remains invariant, so conformally invariant quantities may depend on $\tilde{w}$ only. Note that $I$ is positive apart from eventual boundaries (typically, $I$ may vanish in the asymptotic region and/or at singularities). One may transform to a conformal frame with $I=1$, solve all EOM and then perform the inverse transformation. Thus, it is sufficient to solve the classical EOM for $U=0$,
\begin{gather}
  \extd X + \tilde{X}^-\tilde{e}^+ - \tilde{X}^+\tilde{e}^- = 0 \label{eq:gPSMeom3.1}\ , \\
  (\extd\pm\tilde{\om}) \tilde{X}^\pm \mp \tilde{e}^\pm \tilde{V}(X,B) = 0 \label{eq:gPSMeom3.2}\ ,\\
  (\extd\pm\tilde{\om}) \tilde{e}^\pm = 0 \label{eq:gPSMeom4.2}\ , 
\end{gather}
which is what we are going to do now. Note that the equation containing $\extd\tilde{\om}$ is redundant, while the equations from the Maxwell sector may be treated as in section \ref{se:2.2}; therefore, they have not been displayed.

Let us start with an assumption: $\tilde{X}^{+}\neq 0$ for a given patch.\footnote{To get some physical intuition as to what this condition could mean: the quantities $X^a$, which are the Lagrange multipliers for torsion, can be expressed as directional derivatives of the dilaton field by virtue of \eqref{eq:gPSMeom3.1} (e.g.~in the second order formulation a term of the form $X^aX_a$ corresponds to $(\nabla X)^2$). For those who are familiar with the Newman-Penrose formalism: for spherically reduced gravity the quantities $X^a$ correspond to the expansion spin coefficients $\rho$ and $\rho^\prime$ (both are real).} If it vanishes a (Killing) horizon is encountered and one can repeat the calculation below with indices $+$ and $-$ swapped everywhere. If both vanish in an open region by virtue of \eqref{eq:gPSMeom3.1} a constant dilaton vacuum emerges, which will be addressed separately below. If both vanish on isolated points the Killing horizon bifurcates there and a more elaborate discussion is needed \cite{Klosch:1996qv}. The patch implied by this assumption is a ``basic Eddington-Finkelstein patch'', i.e., a patch with a conformal diagram which, roughly speaking, extends over half of the bifurcate Killing horizon and exhibits a coordinate singularity on the other half. 
In such a patch one may redefine $\tilde{e}^{+}=\tilde{X}^{+} Z$ with a new 1-form $Z$. Then \eqref{eq:gPSMeom3.1} implies $\tilde{e}^{-}=\extd X/\tilde{X}^{+}+\tilde{X}^{-}Z$ and the volume form reads $\tilde{\epsilon}=\tilde{e}^{+}\wedge \tilde{e}^{-}=Z\wedge \extd X$. The $+$ component of \eqref{eq:gPSMeom3.2} yields for the connection $\tilde{\omega}=-\extd \tilde{X}^{+}/\tilde{X}^{+}+Z\tilde{V}(X,B)$. One of the torsion conditions \eqref{eq:gPSMeom4.2} then leads to $\extd Z=0$, i.e., $Z$ is closed. Locally, it is also exact: $Z=\extd u$. It is emphasized that, besides the two Casimir functions, this is the only integration needed! After these elementary steps one obtains already the conformally transformed line element in Eddington-Finkelstein (EF) gauge 
\begin{equation}
  \label{eq:lieelement}
  \extd \tilde{s}^2=2\tilde{e}^{+}\tilde{e}^{-}=2\extd u\,\extd X + 2\tilde{X}^{+}\tilde{X}^{-}\extd u^2\,,
\end{equation}
which nicely demonstrates the power of the first order formalism. In the final step the combination $\tilde{X}^+\tilde{X}^-$ has to be expressed as a function of $X$. This is possible by noting that the linear combination $\tilde{X}^+\times$[\eqref{eq:gPSMeom3.2} with $-$ index] + $\tilde{X}^-\times$[\eqref{eq:gPSMeom3.2} with $+$ index] together with \eqref{eq:gPSMeom3.1} establishes a conservation equation, 
\begin{equation}
  \label{eq:conservation}
  \extd{(\tilde{X}^+\tilde{X}^-)} + \tilde{V}(X,B) \extd X = \extd{(\tilde{X}^+\tilde{X}^- + \tilde{w}(X,B))} = 0\,.
\end{equation}
Thus, there is always a conserved quantity ($\extd \mathcal{C}^{(g)}=0$), which in the original conformal frame reads
\begin{equation}
  \label{eq:c}
  \mathcal{C}^{(g)}=X^+X^-I(X)+\tilde{w}(X,B)\,,
\end{equation}
where the definitions \eqref{eq:wI} have been inserted. It should be noted that the two free integration constants inherent to the definitions \eqref{eq:wI} may be absorbed by rescalings and shifts of $\mathcal{C}^{(g)}$, respectively. Therefore, any mass definition based upon the conserved quantity $\mathcal{C}^{(g)}$ is incomplete without fixing the scale and the ground state geometry.\footnote{This has been clarified for a large class of dilaton gravity models in ref.~\cite{Liebl:1997ti}. Appendix A of ref.~\cite{Grumiller:2004wi} provides a generalization to arbitrary dilaton gravity models.} 
The classical solutions are labelled by this mass. Finally, one has to transform back to the original conformal frame (the relevant conformal factor reads $\Om=I(X)$). The line element \eqref{eq:lieelement} by virtue of \eqref{eq:c} may be written as
\begin{equation}
  \label{eq:EF}
  \extd s^2 =2I(X)\diff u\,\diff X - 2I(X)(\tilde{w}(X,B)- \mathcal{C}^{(g)})\diff u^2\,. 
\end{equation}
Evidently there is always a Killing vector $K\cdot\partial=\partial/\partial u$ with associated Killing norm $K\cdot K=-2I(\tilde{w}-\mathcal{C}^{(g)})$. Since $I\neq 0$ Killing horizons are encountered at $X=X_h$ where $X_h$ is a solution of
\begin{equation}
  \label{eq:horizon}
  \tilde{w}(X_h,B) - \mathcal{C}^{(g)}=0\,.
\end{equation} 
It is recalled that \eqref{eq:EF} is valid in a basic EF patch, e.g., an outgoing one. By redoing the derivation above, but starting from the assumption $X^-\neq 0$ one may obtain an ingoing EF patch. Global issues will be addressed specifically for the ESBH and the ESNS in section \ref{se:3.2.1}.

For sake of completeness it should be mentioned that in addition to the family of solutions, labelled by $M$ and $Q$, isolated solutions may exist, so-called constant dilaton vacua, which have to obey $X=X_{\rm CDV}=\rm const.$ with $V(X_{CDV},B) = 0$. The rank of the Poisson tensor \eqref{eq:poissontensor} vanishes on these solutions. The corresponding geometry has constant curvature, i.e., only Minkowski, Rindler or $(A)dS_2$ are possible spacetimes for constant dilaton vacua. 

\section{The action}\label{se:3}

\subsection{Statement of the main result} \label{se:3.1}

In this paper it will be proven that
the first order Maxwell-dilaton gravity action\footnote{The notation is explained in section \ref{se:2}. Here is a brief summary: the 2-forms $T^\pm=(\extd\pm\om)\wedge e^\pm$, $R=\extd\om$ and $F=\extd A$ are torsion, curvature, and abelian field strength, respectively and depend on the gauge field 1-forms $e^\pm$ (``Zweibein''), $\omega$ (``spin connection'') and $A$ (``Maxwell field''). The scalar fields $X_\pm$, $\Phi$ and $B$ are Lagrange multipliers for these 2-forms and appear also in the potential, the last term in \eqref{eq:solutionofESBH}, which is multiplied by the volume 2-form $\epsilon=e^+\wedge e^-$.}
\begin{equation}
  \label{eq:solutionofESBH}
S_{ESBH}=\int_{{\mathcal M}_2}  \left[X_aT^a+\Phi R + B F + \epsilon \left(X^+X^-U(\Phi)+V(\Phi)\right)\right]\,,
\end{equation}
with potentials $U,V$ to be defined below,
describes the ESBH as well as the ESNS, i.e., on-shell the metric $g_{\mu\nu}=\eta_{ab}\,e^a_\mu\, e^b_\nu$ and the dilaton $X=\exp{(-2\phi)}$ are given by \eqref{eq:dvv2}-\eqref{dvv:dual}. Regarding the latter, the relation
\begin{equation}
  \label{eq:solutionofESBH1} 
  (\Phi-\gamma)^2 = \arcsinh^2{\gamma}
\end{equation}
in conjunction with the definition
\begin{equation}
  \label{eq:solutionofESBH2}
  \gamma:=\frac{X}{B}
\end{equation}
may be used to express the auxiliary dilaton field $\Phi$ in terms of the ``true'' dilaton field $X$ and the auxiliary field $B$. The two branches of the square root function correspond to the ESBH (main branch) and the ESNS (second branch), respectively. Henceforth the notation 
\begin{equation}
  \label{eq:solutionofESBH2.5}
  \Phi_\pm = \gamma \pm \arcsinh{\gamma}
\end{equation}
will be employed, where $+$ refers to the ESBH and $-$ to the ESNS. This applies to all expressions encountered below. If a quantity appears without such  a lower index 
it is the same for ESBH and ESNS.\footnote{In order to avoid confusion with light cone indices, which are also denoted by $\pm$, from now on light cone indices will always appear as upper ones unless stated otherwise, but from the context the meaning should be clear anyhow.} The potentials read
\begin{equation}
  \label{eq:solutionofESBH3}
  V=-2b^2\gamma\,,\quad U_\pm= -\frac{1}{\gamma N_\pm(\gamma)}\,,
\end{equation}
with an irrelevant scale parameter $b\in\mathbb{R}^+$ and
\begin{equation}
  \label{eq:solutionofESBH4}
  N_\pm(\gamma)=1+\frac{2}{\gamma}\left(\frac{1}{\gamma}\pm\sqrt{1+\frac{1}{\gamma^2}}\right)\,.
\end{equation}
Note that $N_+N_-=1$. This completes the definition of all terms appearing in the action \eqref{eq:solutionofESBH}, so what remains to be discussed, apart from the actual proof of equivalence, are the constants of motion. 

On general grounds it is known that two constants of motion exist.\footnote{In the PSM language this statement is particularly easy to derive: the number of constants of motion is given by the dimension of the kernel of the underlying Poisson tensor. For dilaton gravity this amounts to 1, for dilaton gravity coupled to an abelian gauge field it amounts to 2.} One of them is just the field $B$ as can be seen easily from varying \eqref{eq:solutionofESBH} with respect to the gauge field $A$, while the other one is the quantity defined in \eqref{eq:c}. They may be interpreted, respectively, as $U(1)$ charge (cf.~\eqref{eq:casimiru1}),
\begin{equation}
  \label{eq:solutionofESBH7}
  B=\frac{1}{\sqrt{k(k-2)}} e^{-2\phi_0}=:Q\,,
\end{equation}
and mass (cf.~\eqref{eq:c}),
\begin{equation}
  \label{eq:solutionofESBH6}
  {\mathcal C}^{(g)}=-bk=:-2b M\,.
\end{equation}
Note that $Q\geq 0$ and $M\in[1,\infty)$. The restriction on $Q$ to positive values is necessary in order to ensure positivity of $\Phi$ for positive $X$. The restriction on $M$ is not inherent to the model\footnote{See, however, section \ref{se:4} below: Reality of the Hawking temperature \eqref{eq:dvvnew20} also implies $k\geq 2$ without appealing to CFT.} but a consequence of CFT: $k$ may not be smaller than 2 if the string-coupling $\alpha^\prime$ is positive (cf.~\eqref{eq:dvv4}). Scaling and shift ambiguity that exist for any dilaton gravity model are fixed in the next subsection. At the moment $k$ and $\phi_0$ are just some convenient integration constants, but the nomenclature is not accidental: $k$ will turn out to be the level and $\phi_0$ the value of the dilaton at the origin.

As heuristic support for the claim that \eqref{eq:solutionofESBH} is indeed correct one may consider the (singular) limits $\gamma\to 0$ (JT limit) and $\gamma\to\infty$ (Witten BH limit). In the first case for the ESBH branch the effective dilaton $\Phi=2\gamma + {\mathcal O}(\gamma^3)$ immediately reveals the proper potential $V(\Phi)=-b^2\Phi+\dots$ (cf.~\eqref{eq:potentialsJT}). The ESNS branch becomes singular, $U_-\propto 1/\gamma^3$, and is discarded for the time being, while the ESBH branch yields $U_+\propto\gamma$. However, since this potential is multiplied by $X^+X^-$ which is also small in that limit,\footnote{This follows immediately from the EOM \eqref{eq:gPSMeom3.1}: if $\extd X={\mathcal O}(\epsilon)$ and $e^\pm={\mathcal O}(1)$ then $X^\pm={\mathcal O}(\epsilon)$.} it may be dropped to leading order and the result is
\begin{equation}
  \label{eq:JTlimit}
  \left.X^+X^-U\right|_{\gamma\ll 1} = {\mathcal O}(\ga^3)\,,\quad \left.V\right|_{\ga\ll 1} = -b^2\Phi+\mathcal{O}(\ga^3)\,.
\end{equation}
In the second case the limit reads $\Phi_\pm=\gamma(1\pm\mathcal{O}(\ln{\ga}/\ga))$ and the quantities $N_\pm=1\pm \mathcal{O}(1/\ga)$ approach unity. Thus, the ESBH and the ESNS branch coincide, i.e., the Witten BH becomes a self-dual model, and the potentials read
\begin{equation}
  \label{eq:WittenBHlimit}
  \left.X^+X^-U\right|_{\ga\gg 1}\approx-\frac{X^+X^-}{\Phi}\,,\quad \left.V\right|_{\ga\gg 1}\approx-2b^2\Phi\,.
\end{equation}
These limits may be compared with \eqref{eq:potentialsJT} and \eqref{eq:potentialwittenbh}, respectively. However, because both limits are singular this it merely a weak consistency check. 

To prove that \eqref{eq:solutionofESBH} is indeed the correct action one has to study its classical solutions.
It will be shown below that all of them globally coincide with the ones of DVV for any values of $k,\phi_0$ (and $b$) and for both branches $\pm$. On the other hand, for non-negative $Q$ and $M\geq 1$, \eqref{eq:solutionofESBH} does not contain any solution in its spectrum that is not in the DVV family,\footnote{There are some isolated solutions, so-called ``constant dilaton vacua'', $X=X^+=X^-=0$, $B=Q$ which yield $AdS_2$ with curvature scalar $r=-4b^2\extd\ga/\extd\Phi=-2b^2$ for the ESBH branch and a singular result for the ESNS branch, equivalent to the JT limit. This can be understood easily because $V=0$, the necessary condition for a constant dilaton vacuum, implies $\ga=0$, the limit invoked to obtain \eqref{eq:JTlimit}. Therefore, these isolated solutions do not reveal new geometries, but they do yield a new solution for the dilaton, namely $X=0$. \label{fn:CDV}} so it is, up to equivalence transformations, ``the action for the exact string BH''.

\subsection{Proof of equivalence to DVV}\label{se:3.2}

For practically all applications (see sections \ref{se:4} and \ref{se:6} for details) the most interesting part of the construction of the action \eqref{eq:solutionofESBH} is the result itself. That is the reason why the presentation I have chosen turns history around and starts with the result in the previous subsection and proceeds to prove its validity in the current one. The drawback of this is that the reader gets little insight into the actual construction of the action which has been achieved in a bottom-up manner rather than the top-down way presented here. Therefore, it is recalled at this point that the reader interested in a bottom-up construction is invited to consult the appendices. In particular, appendix \ref{app:B.1} recalls the no-go result, including the pivotal feature of dilaton-shift invariance, and provides some first hints how to circumvent it; appendix \ref{app:B.2} demonstrates the crucial idea of ``integrating in'' an abelian $BF$ term in order to promote what otherwise would be a parameter of the action to a constant of motion, namely the constant $\phi_0$.  
This is a vital requirement because otherwise there would be a one-parameter family of actions, labelled by $\phi_0$, and dilaton-shift invariance would not map a solution to another one, but move within that parameter family.
Equipped with the considerations of the appendices one may now reverse-engineer the action from the knowledge of all classical solutions, see section \ref{se:2.3} for details. To follow this reverse-engineering one may essentially read this subsection backwards. Regardless of how the (rather irrelevant) scaling- and shift-ambiguity contained in the functions $I$ and $\tilde{w}$ (cf.~\eqref{eq:wI}) is fixed one obtains {\em uniquely}\footnote{If one weakens the requirements and one considers not just the ESBH/ESNS but a conformal equivalence class (such that all members of this class are related by a conformal transformation with a conformal factor which is regular globally but which may be singular in the asymptotic region $\Phi\to\infty$) then only a certain combination of the two potentials remains unique, namely the function $\tilde{w}$ as defined in \eqref{eq:wI}, calculated below in \eqref{eq:wtilde}. For certain applications the consideration of such equivalence classes may be sufficient. The most convenient representant of this class typically is the one where the transformed potential $\tilde{U}$ vanishes. The corresponding potential $\tilde{V}$ is written below in \eqref{eq:trafoV}.} the potentials $U$ and $V$ as presented in \eqref{eq:solutionofESBH3}. 

\subsubsection{All classical solutions derived from the action}\label{se:3.2.1}

We now proceed to obtain globally all classical solutions derived from the action \eqref{eq:solutionofESBH} and to demonstrate that they coincide with the ones obtained by DVV. 

As a first step one can eliminate the pair $A,B$ and replace the latter by its on-shell value \eqref{eq:solutionofESBH7} in all other EOM. As the potential in \eqref{eq:solutionofESBH} does not depend on $B$ explicitly the abelian field strength vanishes on-shell, $F=0$. Thus, not only the role of $B$ but also the one of $A$ is solely of auxiliary nature. The remaining EOM obtained by variation with respect to $\Phi,X^\pm,\om$ and $e^\pm$ may be solved like in section \ref{se:2.3}. To this end one has to determine first the functions $I_\pm$, $\tilde{w}_\pm$ as defined in \eqref{eq:wI}.

By plugging $U_\pm$ into the left definition \eqref{eq:wI} and exploiting the relation
\begin{equation}
  \label{eq:relation1}
  \frac{\extd\Phi_\pm}{\extd\ga}=1\pm\frac{1}{\sqrt{\ga^2+1}}
\end{equation}
one obtains\footnote{For $I_-$ the multiplicative factor has been chosen to be negative. Consequently, $I_-$ is manifestly negative. The positive side effect of this choice is that one may still consider the change between ESBH and ESNS as a switch of branches of the square root function.}
\begin{equation}
  \label{eq:integratingfactor}
  I_\pm=\frac{1}{2b}\,\frac{1}{1\pm\sqrt{\ga^2+1}}\,,
\end{equation}
where the multiplicative ambiguity from the integration constant inherent to the left definition \eqref{eq:wI} has been fixed by introducing the scale factor $1/(2b)$. It is emphasized that it is completely irrelevant how to fix this factor; the choice in \eqref{eq:integratingfactor} has been made for later convenience.

Now this result and the other potential $V$ may be plugged into the right definition \eqref{eq:wI}. This implies
\begin{equation}
  \label{eq:wtilde}
  \tilde{w}_\pm=\mp b\sqrt{\ga^2+1} + \tilde{w}^0
\end{equation}
It turns out to be helpful to fix the integration constant $\tilde{w}^0$ in \eqref{eq:wtilde} such that
\begin{equation}
  \label{eq:MGS}
  \tilde{w}_\pm=-\frac{1}{2I_\pm}\,.
\end{equation}
This can be achieved with
\begin{equation}
  \label{eq:w0}
  \tilde{w}^0=-b\,.
\end{equation}
Again the choice for \eqref{eq:w0} is merely dictated by convenience. 
Plugging \eqref{eq:integratingfactor} and \eqref{eq:MGS} into the general solution \eqref{eq:EF} promotes it to
\begin{equation}
  \label{eq:line1}
  \extd s^2_\pm= \frac{2\extd u\extd\Phi_\pm}{2b\left(1\pm\sqrt{\ga^2+1}\right)} + \left(1+\frac{\mathcal{C}^{(g)}}{b\left(1\pm\sqrt{\ga^2+1}\right)}\right)\extd u^2\,.
\end{equation}
This is precisely the line element of the ESBH ($+$) and the ESNS ($-$), as will be made explicit in section \ref{se:3.2.2}. 

\FIGURE[t]{
\epsfig{file=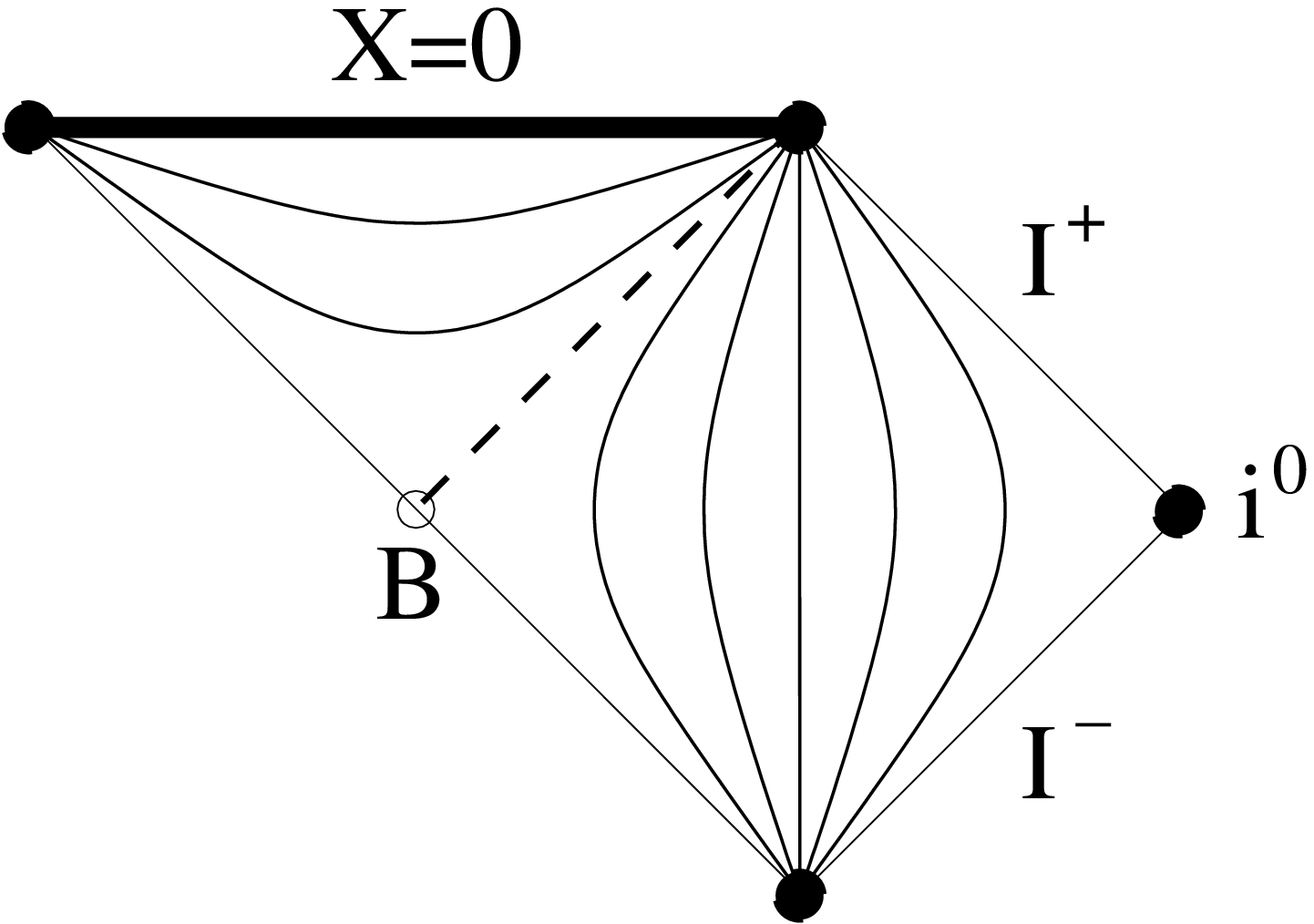,width=0.4\linewidth}\quad\quad\quad\quad\quad
\epsfig{file=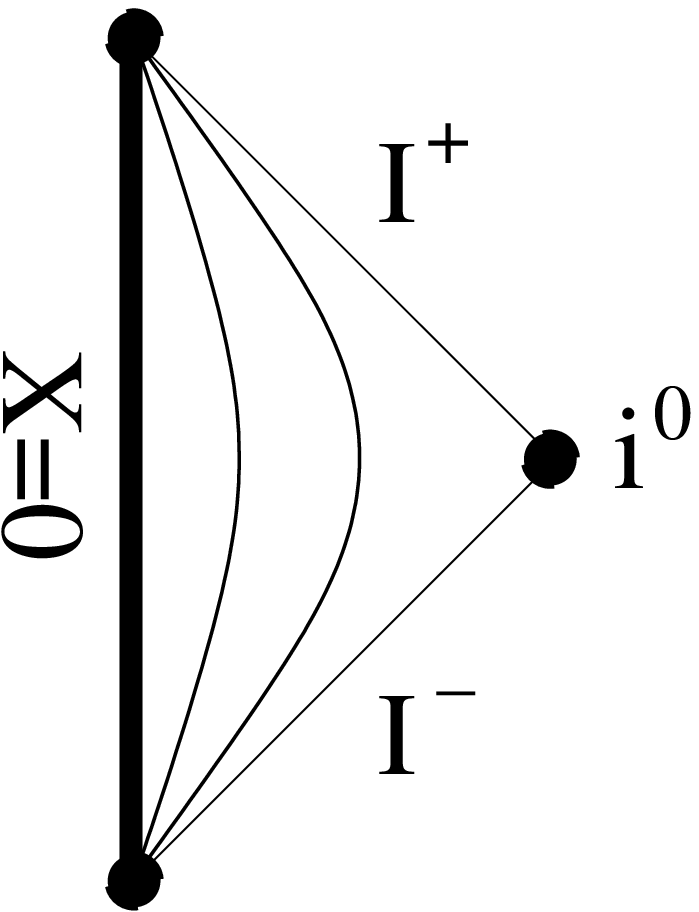,width=0.2\linewidth}
\caption{CP diagrams of EF patches for the ESBH (left) and the ESNS (right).}
\label{fig:CP}
}

The line element \eqref{eq:line1} covers a basic EF patch (see fig.~\ref{fig:CP}; the points 
$i^0,I^+,I^-$ and $B$ denote spatial infinity, future light-like infinity, past light-like infinity and the bifurcation point, 
respectively; the Killing horizon is denoted by the dashed line, the boundary $X=0$, which is regular for the ESBH and singular for the ESNS, by a bold line; curved lines are $X=\rm const.$ hypersurfaces). The global Carter-Penrose (CP) diagram may be obtained by well-known methods \cite{Walker:1970,Klosch:1996qv}: the basic idea is that by simple mirror flips (e.g.~by assuming that $X^-\neq 0$ instead of assuming $X^+\neq 0$ in a given patch) one achieves {\em almost} a universal covering. The only exceptional points\footnote{In the language of general relativity these points are known as bifurcation points $B$, e.g.~the bifurcation 2-sphere in the CP diagram of the Schwarzschild BH.} are those where $X^+=0=X^-$. Around each of these points one may use a Kruskal-like gauge to get an open region containing them. For the ESNS no such complication arises as there is no horizon, while for the ESBH exactly one such point exists for each solution. Consequently, the CP diagram of the ESBH is very similar to the one of the 2D part of the Schwarzschild BH, except that there is no singularity inside the BH region for the ESBH. Although straightforward, rather than performing the construction of the maximally extended space-time as outlined in this paragraph one may consult refs.~\cite{Perry:1993ry,Yi:1993gh}.

\subsubsection{Coordinate transformations to the ESBH}\label{se:3.2.2}

Some simple transformations are invoked in order to bring \eqref{eq:line1} to a more familiar form. With the Casimir $\mathcal{C}^{(g)}$ as parametrized in \eqref{eq:solutionofESBH6}, the definition
\begin{equation}
  \label{eq:wdef}
  w_\pm:=\pm w:=\pm \sqrt{\ga^2+1}\,,
\end{equation}
and the relations
\begin{equation}
  \label{eq:relwphi}
  \frac{\extd\Phi_\pm}{\extd w_\pm}=\sqrt{\frac{w_\pm+1}{w_\pm-1}}\,,
\end{equation}
the line element \eqref{eq:line1} may be written as
\begin{equation}
  \label{eq:line2}
  \extd s^2_\pm= \frac{2\extd u\extd w}{2b\sqrt{w^2-1}} + \left(1- \frac{k}{1\pm w}\right)\extd u^2\,.
\end{equation}
An alternative, even simpler, representation of the line element is provided by the transformation $w=\cosh{(2bR)}$:
\begin{equation}
  \label{eq:line3}
  \extd s^2_\pm=2\extd u\extd R + \xi_\pm(R) \extd u^2
\end{equation}
with the Killing norm
\begin{equation}
  \label{eq:killingnorm}
  \xi_\pm(R)=1-\frac{2M}{1\pm\cosh{(2bR)}}\,.
\end{equation}
This is equivalent to the one by DVV, where $+$ refers to the ESBH and $-$ to its dual solution, the ESNS; this can be seen easily for the ESBH: by virtue of the coordinate transformation\footnote{Note that this last coordinate transformation exhibits the usual coordinate singularity on a Killing horizon $\xi_+(R)=0$ that arises always in the transition from EF gauge to diagonal gauge. Besides this expected singularity there are no further ones for $k>2$. Note that all previous coordinate redefinitions were well-defined globally. An analogous transformation can be applied for the ESNS.} $\extd \tau=\sqrt{1-2/k}\,(\extd u+\extd R/\xi_+(R))$ and $\cosh{(2bR)}=((k-2)\cosh^2{(bx)}+1)$ one obtains with
\begin{equation}
  \label{eq:deff}
  \extd R =  \sqrt{1-2/k}\,f(x)\extd x
\end{equation}
the line element 
\eq{
\extd s^2_+=f^2(x)\extd\tau^2-\extd x^2\,,
}{eq:dvv1rep}
where
\eq{
f(x)=\frac{\tanh{(bx)}}{\sqrt{1-p\tanh^2{(bx)}}}\,.
}{eq:dvv2rep}
This is identical to \eqref{eq:dvv5} with \eqref{eq:dvv2}.

\subsubsection{The dilaton field}\label{se:3.2.3}

At this point it is recalled that it is crucial for the ESBH action to produce both, the dilaton field \eqref{eq:dvv3} and the line element \eqref{eq:dvv1}, \eqref{eq:dvv2}. It is possible, for instance, to construct actions which yield the correct dilaton field but only approximate the appropriate line element (for a concrete example cf.~appendix B.1 in \cite{Grumiller:2002md}). 

Having established that \eqref{eq:solutionofESBH} produces the correct line element, the dilaton field remains to be discussed. The ``true'' dilaton field $X$ may be obtained from \eqref{eq:solutionofESBH2} together with \eqref{eq:solutionofESBH7} in terms of $\gamma$. For the ESBH one may now relate $\gamma$ with the coordinate $x$ by virtue of the transformations \eqref{eq:wdef}-\eqref{eq:deff}:
\begin{equation}
  \label{eq:diltrafo}
  \sqrt{X^2/B^2+1}=\sqrt{\ga^2+1}=w=\cosh{(2bR)}=(k-2)\cosh^2{(bx)}+1
\end{equation}
On-shell this leads to
\begin{equation}
  \label{eq:dilatonresult}
  X=e^{-2\phi_0}\cosh^2{(bx)}\sqrt{(1-2/k)+2/k\cosh^{-2}{(bx)}}\,.
\end{equation}
Undoing the exponential representation implied by \eqref{eq:dilatondilaton} and inserting $p=2/k$ amounts to
\begin{equation}
  \label{eq:dilatonoriginal}
  \phi=\phi_0-\ln{\cosh{(bx)}}-\frac14 \ln{(1-p\tanh^2{(bx)})}\,,
\end{equation}
which is equivalent to \eqref{eq:dvv3}.

\subsection{Discussion and reformulations of the action}\label{se:3.3}

The main result \eqref{eq:solutionofESBH} displays several unexpected features: while in retrospect it may seem obvious that an abelian $BF$ term is capable to circumvent the no-go result of \cite{Grumiller:2002md}, it is slightly surprising that neither $U$ nor $V$ depend explicitly on $B$ if expressed in terms of $\Phi$. This is in contrast to both examples in section \ref{se:2.2} and also in contrast to the Cangemi-Jackiw formulation of Verlinde's first order version of the conformally transformed Witten BH, which motivated the introduction of an abelian $BF$ term for the ESBH in the first place (cf.~appendix \ref{app:B.2}). It suggests strongly that the auxiliary dilaton field $\Phi$, rather than the ``true'' dilaton field $X$, should be taken as primary degree of freedom, in which case the abelian $BF$ term decouples completely and may be integrated out without leaving a trace in the action. Therefore, the constant $\phi_0$ -- or, equivalently, the $U(1)$ charge $B=Q$ -- must not play any physical role in the absence of matter. Indeed, as will be shown in section \ref{se:4}, neither mass, nor temperature, nor entropy depend on it; the same holds for all other quantities of physical interest discussed below, like Killing-norm, curvature, specific heat, evaporation time or free energy.\footnote{It is somewhat ironic that the crucial step to circumvent the no-go result of \cite{Grumiller:2002md} has been the introduction of an abelian $BF$ term which in retrospect, upon proper identification of what should be regarded as primary degree of freedom in the dilaton sector, decouples completely from the theory and may be integrated out again. However, as soon as matter is coupled to the system the value of $\phi_0$ (and hence the field $B$) may be of physical significance -- for instance, $\phi_0$ may play the role of a relative coupling constant between geometric and matter action; this issue is addressed at the end of section \ref{se:matter}.}
Additionally, the strictly monotonous but non-algebraic relation \eqref{eq:solutionofESBH2}-\eqref{eq:solutionofESBH2.5} between the dilaton fields $\Phi$ and $X$ is not something that could have been anticipated a priori. The explicit form of $V$ and $U$ is far less surprising, but the relation between ESBH and ESNS via $N_+N_-=1$ is interesting and maybe deserves a deeper explanation.

A parenthetical remark concerns the use of the first order formulation to derive the action \eqref{eq:solutionofESBH}: as shown in section \ref{se:2.3} and as witnessed by several precedents (for a recent example compare e.g.~\cite{Guralnik:2003we} with \cite{Grumiller:2003ad}) the first order formulation seems to be the most adequate language to describe 2D dilaton gravity. The current paper may also be considered as a demonstration of this assertion. But of course, as both versions of the theory are physically equivalent, somebody might have performed analogous steps to arrive directly at a second order action. The next paragraphs are devoted to the second order formalism.

There are three second order formulations of the action \eqref{eq:solutionofESBH}-\eqref{eq:solutionofESBH4} which may be useful in various contexts, albeit the first one, \eqref{eq:ESBH2ndv1} below, appears to be superior to the other two due to its simplicity and because it invokes directly the dilaton field $\Phi$ rather than $X$. 
We will focus exclusively on the ESBH, but again the results for the ESNS follow straightforwardly from switching to the second branch of the square root function. Eliminating the auxiliary fields $X^a$ and the spin connection $\om$ yields
\begin{equation}
  \label{eq:ESBH2ndv1}
  S_{ESBH}^{(2.1)}=-\frac12 \int_{{\mathcal M}_2}  \extd^2x\sqrt{-g}\left[\Phi \,r - 2 B \,f + U(\Phi) (\nabla\Phi)^2 - 2V(\Phi)\right]\,,
\end{equation}
where $g$ is the determinant of the metric $g_{\mu\nu}$ with respect to which the covariant derivative $\nabla$ is torsion free and metric compatible. The curvature scalar $r$ is multiplied by the auxiliary dilaton $\Phi$ which obeys the relations \eqref{eq:solutionofESBH1}-\eqref{eq:solutionofESBH2.5}. The auxiliary field $B$ on-shell is constant, cf.~\eqref{eq:solutionofESBH7}, while the 2-form field strength is (Hodge) dual to $f$, $\ast F=f$. 
The potentials $U,V$ in \eqref{eq:ESBH2ndv1} are equivalent to the ones in the first order formulation, \eqref{eq:solutionofESBH3} with \eqref{eq:solutionofESBH4}.
Inserting explicitly the dilaton $X$ and using the exponential representation $X=\exp{(-2\phi)}$ leads to the action
\begin{multline}
  \label{eq:ESBH2ndv2}
  S_{ESBH}^{(2.2)}=-\frac12 \int_{{\mathcal M}_2}  \extd^2x\sqrt{-g}\frac{e^{-2\phi}}{B}\Bigg[\left(1 + Be^{2\phi}\arcsinh{\frac{e^{-2\phi}}{B}}\right) \,r - 2 B^2e^{2\phi} \,f \\
-\left(1+\frac{1}{\sqrt{e^{-4\phi}/B^2+1}}\right)^2\frac{B^2e^{4\phi}}{N_+(e^{-2\phi}/B)}\left(\nabla \frac{e^{-2\phi}}{B}\right)^2+4b^2\Bigg]\,,
\end{multline}
Note that variation with respect to the Maxwell-field still results in $B=\rm const.$ on-shell.
 
It may also be of use to transform \eqref{eq:solutionofESBH} to a conformal frame where $\tilde{U}=0$ because some of the physical observables may depend only on the linear combination $\tilde{w}$ in \eqref{eq:wI} which is conformally invariant. It is emphasized that two conformally related theories are inequivalent, in general, especially because the conformal factor necessarily becomes singular for $\Phi\to\infty$. Fixing the multiplicative constant inherent in $\tilde{I}$ such that $\tilde{I}=1/(2b)$, the property $\tilde{I}\tilde{V}=IV=\extd\tilde{w}/\extd\Phi$ implies for the transformed potentials 
\begin{equation}
  \label{eq:trafoV}
  \tilde{V}=-\frac{2b^2\gamma}{1+\sqrt{\ga^2+1}}\,,\quad \tilde{U}=0\,,
\end{equation}
to be inserted into  \eqref{eq:solutionofESBH} or into \eqref{eq:ESBH2ndv1} instead of $V$ and $U$, respectively. Also the transformed version of \eqref{eq:ESBH2ndv2} simplifies considerably:
\begin{multline}
  \label{eq:ESBH2ndv3}
  S_{ESBH}^{(2.3)}=-\frac12 \int_{{\mathcal M}_2}  \extd^2x\sqrt{-g}e^{-2\phi}\Bigg[\left(\frac 1B + e^{2\phi}\arcsinh{\frac{e^{-2\phi}}{B}}\right) \,r - 2 B e^{2\phi} \,f \\
+ \frac{4b^2}{B+\sqrt{e^{-4\phi}+B^2}} \Bigg]\,,
\end{multline}
This may be a convenient starting point for the construction of a Born-Infeld like action: variation with respect to $B$ establishes a non-differential equation in $B$ in terms of $\phi$, $r$ and $f$. Plugging the solution (which need not be unique) back into \eqref{eq:ESBH2ndv3} then produces an action which is highly non-linear in curvature $r$ and field strength $f$. However, such non-linearities are undesirable because thermodynamical discussion (which involves the evaluation of boundary terms), supersymmetrization and quantization are impaired, if not rendered impossible. A comparable action of this type, even without Maxwell field, may be constructed as follows:\footnote{I am grateful to Arkady Tseytlin for providing this argument.} in section 2 of ref.~\cite{Tseytlin:1993df} it was shown that the Witten BH may be transformed into the ESBH with some non-linear field redefinitions of metric and dilaton, containing curvature and derivative terms of the dilaton field non-polynomially. Applying the same field redefinitions to the leading order action (which describes the Witten BH) yields an action for the ESBH which inherits these non-polynomialities. By contrast, the actions \eqref{eq:solutionofESBH}, \eqref{eq:ESBH2ndv1}, \eqref{eq:ESBH2ndv2} and \eqref{eq:ESBH2ndv3} are all linear in curvature and contain no higher derivatives than second ones. This difference is crucial for most of the subsequent applications. In fact, one may consider linearity in curvature as a {\em conditio sine qua non} for a profitable non-perturbative action. 


It has been argued in section \ref{se:3.1} that the weak and strong coupling limits correctly produce the Witten BH and JT model, respectively. As a simple cross-check it is now studied to what extent this holds at the level of solutions, i.e., whether or not curvature as derived e.g.~from \eqref{eq:line3} approaches the appropriate limits. The curvature scalar $r$ is given by (minus) the second derivative of the Killing norm \eqref{eq:killingnorm}:
\begin{equation}
  \label{eq:curvature}
  r_\pm=-\frac{\extd^2\xi_\pm}{\extd R^2}=8b^2M\frac{\cosh^2{(2bR)-2\mp\cosh{(2bR)}}}{\left(1\pm\cosh{(2bR)}\right)^3}
\end{equation}
Nota bene: $r_+$ remains bounded for all $R\in(-\infty,\infty)$ and $r_-$ is singular at $R=0$.
If evaluated at the Killing horizon $r_+$ reduces to
\begin{equation}
  \label{eq:curvaturehorizon}
  \left. r_+\right|_{R=R_{\rm horizon}} = 4b^2\left(1-\frac{3}{2M}\right)\,.
\end{equation}
In the JT limit $R$ becomes very small, $2bR=\eps\ll 1$, and mass goes to $M\to 1$. Curvature then simplifies considerably:
\begin{equation}
  \label{eq:curvatureJT}
  r_+=-2b^2(1+\mathcal{O}(\eps))\,,\quad r_-\propto\eps^{-4}(1+\mathcal{O}(\eps)),.
\end{equation}
As expected the ESBH tends to $AdS_2$ while the ESNS becomes singular. In the Witten BH limit $R$ and $M$ become large, $2bR\gg1$, $M\gg 1$, and curvature reduces to
\begin{equation}
  \label{eq:curvatureWittenBH}
  r_\pm = \pm 16b^2 M e^{-2bR}(1+\mathcal{O}(e^{-2bR}))\,,
\end{equation}
which correctly describes the Witten BH. Self-duality is apparent in \eqref{eq:curvatureWittenBH}. Regarding the global structure it is recalled that for the ESBH the line $X=0$ is regular while for the ESNS it is singular (cf.~fig.~\ref{fig:CP}). However, in the limiting cases these properties change: the Witten BH has the same CP diagram as the ESBH, except that the line $X=0$ is singular; on the other hand, the JT model implies a CP diagram similar to the one of the ESNS, except that the line $X=0$ is regular and space-time is not asymptotically flat; instead it is $AdS_2$, so globally the CP diagram has the form of a vertical strip rather than triangular shape. These discontinuous changes of the causal structure in the limiting cases $k=2$ and $k=\infty$ (or, from \eqref{eq:dvv4}, $\al^\prime=\infty$ and $\al^\prime=0$, respectively) concur with prior observations on the singularity of these limits.

Let us now consider the dual to the JT model in detail. For small $\ga$ one obtains
\begin{equation}
  \label{eq:dualJT}
  \left.N_-\right|_{\ga\ll 1} \approx \frac{\ga^2}{4}\,,\quad \left.\Phi_-\right|_{\ga\ll 1} \approx \frac{\ga^3}{6}\,,
\end{equation}
and consequently the potentials read
\begin{equation}
  \label{eq:dualJTpotentials}
  \left.U_-\right|_{\ga\ll 1} \approx -\frac{2}{3\Phi_-}\,,\quad \left.V\right|_{\ga\ll 1} \approx -2b^2 (6\Phi_-)^{1/3}\,.
\end{equation}
This is a special case of the so-called $ab$-family \cite{Katanaev:1997ni} with $a=2/3$ and $b=-1/3$. Because of $a=b+1$ it is also a Minkowskian ground state model. Somewhat surprisingly, spherical reduction of the 5D Einstein-Hilbert action produces the same model. 
Thus, winding/momentum mode duality connects $AdS_2$ with the 5D Schwarzschild BH.

There are two more points worthwhile to address: It is an unexpected result that mass as given by \eqref{eq:solutionofESBH6} is determined by the level $k$ rather than by the value of the dilaton at the origin, $\phi_0$ -- on the issue of mass cf.~sections \ref{se:4.2} and \ref{se:string}. The introduction of an abelian gauge field has been a crucial input to circumvent the no-go result of ref.~\cite{Grumiller:2002md}. For details on this idea the appendices \ref{app:B.1} and \ref{app:B.2} may be consulted.

\section{Thermodynamical properties}\label{se:4}

Now that a suitable action for the ESBH exists one may employ it to calculate thermodynamical quantities of interest: ADM mass, Hawking temperature and Beken\-stein--Hawking entropy. In each case first a simple and then a more elaborate derivation 
will be provided or at least sketched.
A comparison with previous results \cite{Witten:1991yr,Gibbons:1992rh,Nappi:1992as,Perry:1993ry,Liebl:1997ti,Kazakov:2001pj} will be postponed until section \ref{se:6}.  

\subsection{Mass}\label{se:4.1}

We follow the prescription of the appendix of \cite{Grumiller:2004wi} to obtain the ADM mass which slightly generalizes the procedure presented in \cite{Liebl:1997ti}. The ``zeroth step'' is evident from \eqref{eq:integratingfactor}-\eqref{eq:wtilde} and just amounts to the definition of the functions $I$ and $\tilde{w}$, where already a convenient scale factor has been included by hand in the definition of $I$. The ``first step'', i.e., the identification of the ground state geometry, has been performed in \eqref{eq:w0}, implying a Minkowskian ground state. Such an identification is not just convenient, but necessary if by ``mass'' we mean ``ADM mass''. The asymptotic region to be obtained in the ``second step'' is located at $X\to\infty$ (or alternatively $\ga\to\infty$ or $\Phi\to\infty$); the scaling ambiguity already has been fixed, and \eqref{eq:MGS} implies that mass is measured in units of the asymptotic Killing time. The ADM mass is given by 
\begin{equation}
  \label{eq:ADM}
  M_{\rm ADM} = -\mathcal{C}^{(g)}  = bk = \frac{2b}{p} = \frac{1+2\al^\prime b^2}{\al^\prime b} = 2bM\,.
\end{equation}
In this manner the mass defined in \eqref{eq:solutionofESBH6} is already the ADM mass up to the scale factor introduced by hand in \eqref{eq:integratingfactor}. 

One may derive \eqref{eq:ADM} also by considerations a la Gibbons-Hawking from a boundary term \cite{Gibbons:1977ue}. In the context of 2D dilaton gravity this procedure is described in detail e.g.~in section 5 of \cite{Grumiller:2002nm}. Following it for the ESBH the result is (cf.~(5.10) in  \cite{Grumiller:2002nm})
\begin{equation}
  \label{eq:ADMGH}
  M_{\rm ADM}=\lim_{R\to\infty}\sqrt{\xi(R)}\left(1-\sqrt{\xi(R)}\right)\partial_R \Phi\,,
\end{equation}
where $\xi(R)=\xi_+(R)$ may be read off from \eqref{eq:killingnorm}.
Actually, the least trivial term in \eqref{eq:ADMGH} is the last one: the auxiliary dilaton field $\Phi$ enters here because it is the one which multiplies the Ricci scalar in the action -- and for the calculation of the ADM mass from a boundary term only the contribution $\Phi R$ to the action is of importance (to prevent notationally induced hazard: $R$ refers here to the curvature 2-form, while in the rest of this paragraph it denotes the coordinate introduced in \eqref{eq:line3}). For large values of $R$ the relations
\begin{equation}
  \label{eq:largeR}
  \partial_R\Phi \approx be^{2bR}\,,\quad \sqrt{\xi(R)}\approx 1 - 2 M e^{-2bR}
\end{equation}
imply \eqref{eq:ADM} when plugged into \eqref{eq:ADMGH}. If the action \eqref{eq:solutionofESBH} is multiplied by some overall constant $\kappa$ then also the boundary term and hence the right hand side in \eqref{eq:ADMGH} are multiplied by $\kappa$, so $M_{\rm ADM}\to\kappa M_{\rm ADM}$. One has to bear in mind this simple fact when applying different conventions for the action.

It should be emphasized that the ADM mass depends only on the parameter $b$ and the string-coupling $\al^\prime$. In particular, it does {\em not} depend on $\phi_0$. For the Witten BH mass is usually identified with some function of $\phi_0$ -- but the Witten BH is an exceptional point at the edge of the ESBH family. It is easy to comprehend where this apparent discrepancy comes from: the action \eqref{eq:solutionofESBH} is dilaton shift invariant according to \eqref{eq:dilshift}, while the geometric part of the CGHS action (which describes the Witten BH) is not dilaton shift invariant; rather, only the EOM are. The same considerations apply to the corresponding boundary terms and thus in the former case the mass becomes independent from $\phi_0$ while in the latter it does not. For sake of clarity it pays to plug the limit \eqref{eq:WittenBHlimit} together with the on-shell value of $B$, \eqref{eq:solutionofESBH7}, into the general action \eqref{eq:solutionofESBH}:
\begin{equation}
  \label{eq:scaledwittenaction}
  S^{WBH}=\frac{1}{Q} \int_{{\mathcal M}_2}  \left[\hat{X}_aT^a+X R - \epsilon \left(\frac{\hat{X}^+\hat{X}^-}{X}+2b^2X\right)\right]\,,
\end{equation}
where $\hat{X}^\pm=QX^\pm$. Apart from an overall factor of $1/Q$ this is the geometric part of the CGHS model describing the Witten BH. As we have seen above an overall factor just rescales the ADM mass, which is the reason for the aforementioned absence of $\phi_0$ in \eqref{eq:ADM}. It is stressed that there does not seem to be a ``bottom-up'' way to fix the overall factor (which may neither depend on $k$ nor on $\phi_0$) in front of \eqref{eq:solutionofESBH}, but as its only classical effect is the rescaling of physical units this is of limited relevance.

Because it is nice to work with a dimensionless mass in most of the subsequent considerations the quantity $M\in[1,\infty)$ will be used, so one has to bear in mind the scale factor of $2b$ if one would like to express these results in terms of the ADM mass. It is recalled that the lower boundary value, $M=1$, corresponds to the JT limit, while the upper one, $M\to\infty$, corresponds to the Witten BH limit.

\subsection{Temperature}\label{se:4.2}

Hawking temperature as derived naively from surface gravity 
\begin{equation}
  \label{eq:surfaceg}
  T_H=\frac{1}{4\pi}\left|\frac{\extd \xi}{\extd R}\right|_{R=R_{\rm horizon}}\,,
\end{equation}
 with \eqref{eq:killingnorm} and \eqref{eq:ADM} yields the mass-to-temperature law
\begin{equation}
  \label{eq:dvvnew20}
  T_H=\frac{b}{2\pi}\sqrt{1-\frac{1}{M}}\,.
\end{equation}
Thus, for the Witten BH, $M\to\infty$, one gets $T_H\to b/(2\pi)$, equivalent to the standard behavior found in the literature.
For low masses, $M\to 1$, the temperature $T_H$ vanishes, being consistent with the fact that JT does not exhibit a horizon.

To put the result \eqref{eq:dvvnew20} on firmer grounds an alternative derivation \cite{Christensen:1977jc} is presented which involves a minimally coupled massless scalar field $\scalarf$ as ``carrier'' of the Hawking quanta, the action of which, viz. 
\begin{equation}
  \label{eq:kgaction}
  S_{KG}=\frac12 \int_{\mathcal{M}_2} \extd^2x\sqrt{-g}g^{\mu\nu}\left(\partial_\mu\scalarf\right)\left(\partial_\nu\scalarf\right) \,,
\end{equation}
has to be added to the geometric part \eqref{eq:solutionofESBH}. Vacuum polarization effects determine the semi-classical energy momentum tensor
 \begin{equation}
\label{ehT}
T_{\mu \nu }=\frac{2}{\sqrt{-g}}\frac{\delta W}{\delta g^{\mu \nu }}\, ,
\end{equation} 
where \( W \) is the one-loop effective action for the scalar field on the
classical background manifold with metric \( g_{\mu \nu } \) given by \eqref{eq:line3}, \eqref{eq:killingnorm}.
The trace anomaly\footnote{The notation of section 6 of ref.~\cite{Grumiller:2002nm} is used, with the important exception that the Ricci scalar there, $R$, corresponds to $-r$ in the present work.} 
\begin{equation}
\label{ehmca}
T_{\mu }^{\mu }=-\frac{1}{24\pi}r
\end{equation}
together with the conservation equation
\begin{equation}
\label{ehmcon}
\nabla ^{\mu }T_{\mu \nu }=0\, ,
\end{equation}
then implies a non-vanishing flux component of the energy-momentum tensor. For a review on this method in the context of 2D dilaton gravity cf.~ref.~\cite{Kummer:1999zy}. In conformal gauge (all $\pm$ indices in this subsection refer to the light cone)
\begin{equation}
\label{ehmc2}
\extd s^2=2e^{2\rho }\extd x^{+}\extd x^{-}\, ,\qquad \rho =\frac{1}{2}\ln{\xi}=\frac 12 \ln{\left(1-\frac{2M}{1+\cosh{(2bR)}}\right)}\, ,
\end{equation}
the conservation equation \eqref{ehmcon} yields
\begin{equation}
\label{ehmc-} 
\partial _{+}T_{--}+\partial _{-}T_{+-}-2(\partial _{-}\rho )T_{+-}=0\, .
\end{equation}
Taking into account staticity, the expression for the curvature scalar, $r=-\xi^{\prime\prime}$ (prime denotes differentiation with respect to $R$), as well as $T^{\mu}_{\mu }= 2 e^{-2\rho} T_{+-}$, establishes for the flux component
 \begin{equation}
\label{ehmTl}
T_{--}=\frac{1}{96\pi }\left[ 2\xi \xi^{\prime\prime} -(\xi^\prime )^{2}\right] + t_{0}\, ,
\end{equation}
where $t_0$ is an integration constant. It is fixed by the (Unruh) requirement\footnote{This condition ensures sufficient regularity of the energy-momentum tensor at the Killing horizon in global (Kruskal) coordinates. Other choices select different vacua, e.g.~$t_0=0$ produces Boulware.}
\begin{equation}
\label{ehmUvac}
\left. T_{--}\right|_{R=R_{\rm horizon}}=0\,.
\end{equation}
Consequently, the asymptotic flux is given by
\begin{equation}
  \label{eq:asyflux}
  T_{--}^{\rm asymptotic} = t_0 = \left. \frac{1}{96\pi} (\xi^\prime)^2\right|_{R=R_{\rm horizon}}\,.
\end{equation}
By virtue of the 2D Stefan-Boltzmann law,
\begin{equation}
T_{--}^{\rm asymptotic}=\frac{\pi}{6}T^2_H\,, 
\label{eq:2dsb}
\end{equation}
one then derives \eqref{eq:surfaceg}, so this method gives the same result for the Hawking temperature as the purely geometric one above.

\subsection{Entropy}\label{se:4.3}

Simple thermodynamic considerations establish that entropy $S$ is proportional to the dilaton field evaluated on the Killing horizon \cite{Gegenberg:1995pv}, where ``the dilaton field'' again refers to the factor multiplying the Ricci scalar in the action. Thus, for the ESBH one has to evaluate $\Phi$ at the Killing horizon. With the same convention for the Boltzmann constant as in \cite{Gegenberg:1995pv} the result is
\begin{equation}
  \label{eq:entropy}
  S=\left.2\pi\Phi\right|_{2M=1+\sqrt{\ga^2+1}}=4\pi\left(\sqrt{M(M-1)}+\frac12\arcsinh{(2\sqrt{M(M-1)})}\right)\,.
\end{equation}
This may be understood most easily from $\extd S=\extd M_{\rm ADM}/T_H=4\pi\extd M/\sqrt{1-1/M}$, which upon integration yields
\begin{equation}
  \label{eq:intentropy}
  S=S_0+4\pi\left(\sqrt{M(M-1)}+\arctanh{\sqrt{1-1/M}}\right)\,.
\end{equation}
Setting $S_0=0$ and using simple trigonometric identities for hyperbolic functions, \eqref{eq:intentropy} is seen to be equivalent to \eqref{eq:entropy}. In the Witten BH limit it simplifies to
\begin{equation}
  \label{eq:entropywittenBH}
  S|_{M\gg 1} = 4\pi M\left(1+\mathcal{O}(\ln{(M)}/M)\right) \approx S_{\rm LO} = 4\pi M\,,
\end{equation}
while for the JT limit one gets
\begin{equation}
  \label{eq:entropyJT}
  S|_{M=1+\eps} = 8\pi\sqrt{\eps}\left(1+\mathcal{O}(\eps)\right)\,. 
\end{equation}
Therefore, as might have been anticipated, for $k\to 2$ entropy vanishes. For $k\to\infty$ it is worthwhile to display also the next to leading order term:
\begin{equation}
  \label{eq:lnentropy}
  S|_{M\gg 1} = S_{\rm LO} + 2\pi \ln{S_{\rm LO}} +\mathcal{O}(1)
\end{equation}
Hence, only in the weak coupling regime ($\al^\prime\ll 1$) the expected \cite{Fiola:1994ir,Zaslavsky:1996dg} 
qualitative behavior of entropy \eqref{eq:lnentropy} is recovered, while in the strong coupling regime ($\al^\prime\gg 1$) no logarithmic corrections to entropy do arise.

A different derivation of entropy is provided by counting of microstates with CFT methods using the Cardy formula for the asymptotic density of states (cf.~e.g.~\cite{Strominger:1998yg}). 
We will sketch here a more recent study \cite{Carlip:2004mn} tailor made for 2D dilaton gravity. The main feature of this approach is the imposition of a ``stretched horizon'', i.e., a surface which is ``almost null'', the ``almost'' being parametrized by a small parameter $\epsilon$. Translated to our notation this implies $X^+X^-=\mathcal{O}(\epsilon)\ll 1$. A Hamiltonian analysis is then performed with the boundary condition that a (stretched) horizon must exist. The constraints generate a Virasoro algebra 
with central charge $c=\mathcal{O}(\epsilon)$ which vanishes in the limit $\epsilon\to 0$. Thus, at first glance the Cardy formula\footnote{In \cite{Carlip:2004mn} it has been assumed that the lowest Eigenvalue of $L_0$, denoted by $\Delta_g$, vanishes. One can weaken this requirement and assume that it scales with $\mathcal{O}(\epsilon^2)$ and thus is small as compared to $c$ which scales with $\mathcal{O}(\epsilon)$. Therefore, instead of $c/6-4\Delta_g$ to leading order only $c/6$ is present in \eqref{eq:cardy}. Similarly, the term $\De-c/24$ to leading order is just given by $\De$. Note that there is a relative factor of $1/2$ as compared to \cite{Carlip:2004mn} because the action \eqref{eq:ESBH2ndv1} contains a relative factor of $1/2$ as compared to (4) in \cite{Carlip:2004mn} for $16\pi G=1$.} \cite{Cardy:1986ie} 
\begin{equation}
  \label{eq:cardy}
  S=\pi\sqrt{\frac{c}{6}\, \De}\,,
\end{equation}
where $\De$ is the Eigenvalue of the Virasoro operator $L_0$ for which entropy $S$ is being calculated, appears to produce a vanishing entropy. However, $\De$ turns out to be proportional to $1/\epsilon$. To be more concrete, the results of \cite{Carlip:2004mn} are\footnote{Various approaches yield different results for $c$ and $\De$. However, in their product these ambiguities always seem to disappear.}
\begin{equation}
  \label{eq:carlip}
  c=24\pi\epsilon  \,\Phi_h\,,\quad \De=\frac{\Phi_h}{\pi\epsilon}\left(1+\mathcal{O}(\epsilon)\right)\,,
\end{equation}
where $\Phi_h$ is the dilaton $\Phi$ restricted to the Killing horizon. Inserting \eqref{eq:carlip} into \eqref{eq:cardy} confirms \eqref{eq:entropy}. It is emphasized that in all approaches the specific form of $U$ and $V$ is essentially irrelevant; only the $\Phi R$ term in the action matters. 

\section{Conclusions and generalizations}\label{se:6}

The main result of this work is the action for the ESBH/ESNS, \eqref{eq:solutionofESBH}-\eqref{eq:solutionofESBH4}. It may be considered as a non-perturbative generalization of the geometric part of the CGHS model \cite{Callan:1992rs} valid for all values of the string-coupling $\al^\prime$, while retaining the pivotal property of linearity in curvature. It is worthwhile mentioning that the potentials $U,V$ in \eqref{eq:solutionofESBH3} are unique.
The form of \eqref{eq:solutionofESBH} suggests to take the auxiliary dilaton $\Phi$ more seriously and to treat it as a ``genuine'' dilaton field. In that case the $BF$ term decouples and may be eliminated. Such an action (or, alternatively, its second order cousin \eqref{eq:ESBH2ndv1}) may be a useful starting point for adding matter degrees of freedom, supersymmetrization, quantization, etc. Subsequently a few possible applications are pointed out and some speculations are presented, although the list by no means is meant to be exhaustive. 

\subsection{Supplementary thermodynamical considerations} 
 
Comparing the results of section \ref{se:4} with available ones in the literature \cite{Witten:1991yr,Gibbons:1992rh,Nappi:1992as,Perry:1993ry,Liebl:1997ti,Kazakov:2001pj} almost exclusively disagreement is found. This should not come as a shock, because the derivation of the ADM mass and of the entropy crucially depends on the knowledge of the correct $\Phi R$ term in the action, hitherto unavailable. In most previous derivations it was assumed explicitly or implicitly that the corresponding term in the action reads $X R$, with $X$ as given by \eqref{eq:solutionofESBH2} off-shell and by \eqref{eq:dilatonresult} on-shell.
Only regarding Hawking temperature there is some agreement:\footnote{Although there is no universal agreement among the previous literature on that issue. For instance, \cite{Dijkgraaf:1992ba} gets (translated to our notation) $T_H\propto \sqrt{M-1}$, while \cite{Perry:1993ry} obtains $T_H\propto 1/\sqrt{M}$.} the result \eqref{eq:dvvnew20} coincides with equation (3.12) in  ref.~\cite{Kazakov:2001pj} and with equation (2.10) in ref.~\cite{Yi:1993gh}. This is to be expected because for the derivation of Hawking temperature via surface gravity only knowledge about the metric is needed, but not about the action. So the first obvious application of the action \eqref{eq:solutionofESBH} was a reliable derivation of various thermodynamical quantities, as performed in section \ref{se:4}, which finally clarified what is the ADM mass \eqref{eq:ADM}, mass-to-temperature law \eqref{eq:dvvnew20} and entropy \eqref{eq:entropy} of the ESBH.

It is worthwhile mentioning that the ADM mass need not constitute the most natural mass definition. It is beneficial if the ground state geometry is Minkowski space. But if, say, a BPS solution exists it is often more convenient to shift the mass such that $M({\rm BPS})=0$ and masses of non-BPS solutions are measured with respect to it.\footnote{E.g.~for the Reissner Nordstr\"om BH the ADM mass of the extremal solution is given by $M_{ADM}=|Q|$, so shifting to $M:= M_{ADM}-|Q|$ yields $M=0$ for extremal solutions. For $M\gg Q$ the difference between the two definitions is negligible.} Also the ESBH has a ground state geometry which is not Minkowski space,\footnote{Only analytic continuation to $k<2$ may lead to Minkowski space, which emerges from the limit $k\to 0$. Pushing this further to negative $k$ yields ${\mathcal C}^{(g)}<0$. In that case there are Killing horizons for the ``ESNS'', but not for the ``ESBH''. 
The range $k\in(2,\infty)$ corresponds to $D\in(-\infty,26)$, while $k\in[0,2)$ implies $D\in[29,\infty)$ and $k\in(-\infty,0]$ leads to $D\in(26,29]$. Clearly, $k=2$ (or $D=26$) is an exceptional case. Amusingly, the Minkowski space solution formally requires $D=29$. \label{fn:ksmall}} namely $AdS_2$ for $M=1$. The mass definition
\begin{equation}
  \label{eq:JTmass}
  M_{AdS}:=M_{\rm ADM}-2b=2b(M-1)
\end{equation}
yields $M_{AdS}=0$ for the ground state geometry. In the Witten BH limit the difference between $M_{AdS}$ and $M_{\rm ADM}$ is negligible. For certain applications the ``$AdS$-mass'' \eqref{eq:JTmass} may be more appropriate than the ADM mass \eqref{eq:ADM}. From \eqref{eq:ADMGH} it may be derived easily that for the ESNS the ADM mass is negative because the second term in the right equation of \eqref{eq:largeR} changes sign. This implies a mass gap of $4b$ between the ESBH and the ESNS spectra.

The naively defined specific heat of BHs sometimes exhibits remarkable behavior -- for instance, for the Schwarzschild BH it is negative. Therefore,
it is worthwhile to consider the specific heat for the ESBH,
\begin{equation}
  \label{eq:cs}
  C_s:=\frac{\extd M_{\rm ADM}}{\extd T_H} = \frac{16\pi^2}{b}\,M^2\,T_H\,.
\end{equation}
So the ESBH behaves like an electron gas at low temperature ($T_H\to 0$, $M\to 1$) with Sommerfeld constant $\gamma=16\pi^2/b$. Of particular relevance is the large mass limit, because the uncorrected Witten BH has a vanishing inverse specific heat and thus a finite result is a non-trivial effect:
\begin{equation}
  \label{eq:cswbh}
  \left.C_s\right|_{M\gg 1} =  \frac{2\pi}{b^2} M_{\rm ADM}^2 \left(1+\mathcal{O}(1/M_{\rm ADM})\right)
\end{equation}
It is positive and proportional to the mass squared. Strikingly, up to a numerical factor of $24\pi$ this is precisely the result of \cite{Grumiller:2003mc} obtained by completely different methods which apply to the quantum corrected Witten BH only. In this fashion leading order quantum corrections of the field theoretic approach in \cite{Grumiller:2003mc} qualitatively reveal the same behavior as the stringy corrections implicit in the ESBH.

The Hawking evaporation according to \eqref{eq:dvvnew20} implies a loss of mass and thus the ESBH evolves to another ESBH with lower value of $k$; once $k=2$ is reached Hawking radiation stops, suggesting $AdS_2$ as endpoint of Hawking evaporation. From a string theoretical point of view it appears to be difficult to interpret what an evaporation to a lower level $k$ means -- whether this is a defect of the description of the ESBH via the action \eqref{eq:solutionofESBH} or a new feature predicted by it.\footnote{A possibility to avoid that temperature (and hence the level $k$) is a fixed quantity rather than a free parameter (as required for evaporation) has been addressed below equation (3.12) in \cite{Kazakov:2001pj}: one may assume that temperature is changed by varying the number of extra massless ``matter'' fields that can be added to the system, thereby changing the effective central charge. See also section \ref{se:string}. \label{fn:k}} But as in the case of the CGHS model \cite{Callan:1992rs} one may ``forget'' about the origin of the action and treat it as a model on its own. Then, there is absolutely no interpretational problem with evaporation to lower $k$.
The time interval $\Delta t$ to evaporate from an initial mass $M$ to some final mass $M_0$ according to
\begin{equation}
  \label{eq:evaptime}
  \frac{\extd M}{\extd t} = -\frac{\pi}{6}T_H^2
\end{equation}
upon integration yields
\begin{equation}
  \label{eq:time}
  \Delta t = \frac{24\pi}{b^2}\left(M-M_0+\ln{\frac{M-1}{M_0-1}}\right)\,.
\end{equation}
Therefore, evaporation to $AdS_2$ takes infinitely long, concurrent with its role as ground state geometry.
For large masses the result \eqref{eq:time} essentially coincides with the one derived in \cite{Grumiller:2003mc}. 
On a more speculative side note, it may be rewarding to consider the exceptional solutions mentioned in footnote \ref{fn:CDV}, viz., the constant dilaton vacua, as possible end points of the evaporation process. Their geometry is equivalent to the one of the ground state, $AdS_2$, but the dilaton $X$ vanishes identically. It has been argued in a different context that at the end point of Hawking evaporation a phase transition to a constant dilaton vacuum may occur \cite{Grumiller:2003hq} and it would be excellent to see this falsified or confirmed in a string theoretical derivation.

It could be of interest to determine other quantities, like the free energy $F$. One can derive from it e.g.~the Euclidean action $I=F/T_H$ and the partition function $Z=\exp{(-I)}$. Taking into account that $AdS_2$ is the ground state geometry the definition 
\begin{equation}
  \label{eq:freeenergydef}
  F:=M_{AdS}-T_H S=M_{\rm ADM}-2b-T_H S
\end{equation} 
seems to be preferred. Plugging in the results for mass \eqref{eq:ADM}, temperature \eqref{eq:dvvnew20} and entropy \eqref{eq:entropy} yields
\begin{equation}
  \label{eq:freeenergy}
  F=-b \sqrt{1-\frac{1}{M}}\,\arcsinh{\left(2\sqrt{M(M-1)}\right)}\,.
\end{equation}
Therefore, $F$ is negative apart from the limiting case $M=1$ where $F$ is zero. There is no extremum of $F$ in the range $M\in[1,\infty)$. It might be worthwhile to perform a more elaborate analysis of free energy, e.g.~in analogy to \cite{Kazakov:2001pj}. 
Finally, it could be an interesting exercise to put the ESBH into a cavity of finite size in order to achieve an equilibrium and to carry out a thermodynamical study of the combined system ESBH plus radiation. For recent reviews on BH thermodynamics cf.~e.g.~\cite{Wald:1999vt} 
and refs.~therein.

\subsection{Applications in 2D string theory}\label{se:string}

It is natural to inquire about implications for 2D string theory. First, the role of $k$ as a constant of motion is assessed critically, next the absence of higher derivative terms in the action is highlighted and contrasted with perturbative results, and finally a list of miscellaneous remarks and speculations is presented.

In string theory the level $k$ typically is a fixed quantity, 
while in the theory constructed in the present work it emerges as a constant of motion, essentially the ADM mass. This apparent discrepancy urgently asks for some explanation. Therefore, let us first try to reinterpret $k$ as a parameter of the action: reverting the arguments in appendix \ref{app:B.2} in general it is possible to integrate out certain fields replacing them by their on-shell values. For instance, in the paragraph containing \eqref{eq:app1}-\eqref{eq:app5} a constant of motion stemming from an abelian $BF$-term is converted into a parameter of the action (and vice versa). However, this particular manoeuvre is impossible for the constant of motion corresponding to the mass. So we have to live with the fact that $k$ is not fixed in the action \eqref{eq:solutionofESBH} or one of its reformulations. I will now try to argue that one should not only accept this conclusion but embrace it. There is actually a physical reason why $k$ defines the mass: in the presence of matter the conservation equation $\extd {\mathcal C}^{(g)}=0$ acquires a matter contribution,
\begin{equation}
  \label{eq:consmatter}
  \extd {\mathcal C}^{(g)}+W^{(m)}=0\,,
\end{equation}
where $W^{(m)}=\extd{\mathcal C}^{(m)}$ is an exact 1-form defined by the energy-momentum tensor (cf.~section 5 of \cite{Grumiller:2002nm} or ref.~\cite{Kummer:1995qv}). In a nutshell, the addition of matter deforms the total mass which now consists of a geometric and a matter part, ${\mathcal C}^{(g)}$ and ${\mathcal C}^{(m)}$, respectively. Coming back to the ESBH, the interpretation of $k$ as mass according to the preceding discussion implies that the addition of matter should ``deform'' $k$. But this is precisely what happens: adding matter will in general change the central charge and hence the level $k$. Thus, from an intrinsically 2D dilaton gravity point of view the interpretation of $k$ as mass is not only possible but favoured (see also footnote \ref{fn:k}).

The absence of higher order derivatives in \eqref{eq:solutionofESBH} has been a crucial ingredient for its construction and especially for thermodynamical applications, but it may be difficult to comprehend from a perturbative point of view; after all, an expansion in powers of $\alpha^\prime$ yields arbitrary powers in curvature in the $\beta$ functions and consequently also in corresponding effective actions \cite{Callan:1985ia}. So how is it possible that non-perturbatively curvature appears only linearly? To get some insight, consider a simple model \eqref{eq:GDT} with $U=0$ and $V\propto X^n$ with $n\neq 0,1$. On-shell, the equation $r\propto X^{n-1}$ modulo branch ambiguities allows to eliminate the dilaton field thus obtaining an action non-linear in curvature, with a Lagrangean proportional to $r^{n/(n-1)}$.
If $V$ is not a monomial but a more complicated function similar considerations apply and one obtains an action which may contain an arbitrary Laurent series in $r$. Knowing just (some terms of) the Laurent series, it may be difficult to induce the simpler non-perturbative expression linear in $r$ -- but if the latter is available one may deduce the perturbative results, although it will be a somewhat tedious task. By analogy,\footnote{It is really just an analogy but its message hopefully is transparent: in 2D dilaton gravity various formulations of the same model may exist and equivalence between two seemingly different models with different field content may not be easy to spot just by looking at the actions. The safest check is a global comparison of all solutions (a local comparison is inadequate since all 2D geometries are locally conformally flat). If they coincide the models are classically equivalent. The procedure in the text above eliminates the dilaton field which is not desirable for comparison with results from string theory, but if one follows it from \eqref{eq:solutionofESBH2.5} and \eqref{eq:solutionofESBH3} it is clear that arbitrary powers in $r$ will arise.} it may be difficult to induce from 3- or 4-loop results the non-perturbative action \eqref{eq:solutionofESBH}, but it should be doable to derive these perturbative results from \eqref{eq:solutionofESBH}. 
Because it has been shown in the present work that all classical solutions of the action \eqref{eq:solutionofESBH}-\eqref{eq:solutionofESBH4} coincide globally (!) with the ESBH/ESNS by DVV, it is as reliable as the ESBH itself. So if there were reasons to doubt the validity of the latter of course also the former is obsolete; on the other hand, if one trusts the ESBH -- the working hypothesis of the current work -- one may equally trust \eqref{eq:solutionofESBH}.

Here are some additional loose ends:
\blist
\item It would be gratifying to get a better understanding of the (winding/momentum mode) duality between $AdS_2$ (cf.~\eqref{eq:potentialsJT}) and Schwarzschild in 5D (cf.~\eqref{eq:dualJTpotentials}). To this end one may take advantage of the strong coupling limit $\alpha^\prime\to\infty$.
\item From \eqref{eq:curvaturehorizon} it is evident that $k=2M=3$ is special insofar as curvature vanishes at the horizon; for $k>3$ ($k<3$) it is positive (negative). Incidentally, in the ${\rm \it SL}(2)/U(1)$ CFT the same value of $k$ separates two regions: the CFT exhibits a normalizable zero mode which for $k\leq 3$ becomes non-normalizable (cf.~section 7.3 in ref.~\cite{Karczmarek:2004bw}). This may be either coincidence or of importance for recent discussions on (non)existence of BHs in 2D string theory \cite{Martinec:2004qt,Friess:2004tq,Karczmarek:2004bw}. 
\item The construction of a Born-Infeld like action in terms of dilaton, metric and Maxwell field as outlined below \eqref{eq:ESBH2ndv3} and a comparison with the non-linear action only in terms of dilaton and metric described there may be of interest; however, for such an action the caveats mentioned in the same paragraph apply and limit its pertinence. 
\item A connection with the 3D charged black strings of Horne and Horowitz \cite{Horne:1991gn} has been spelled out in \cite{Yi:1993gh}: dimensional reduction leads to a 2D model exhibiting one PPDOF, which does not describe solely the ESBH. But its static solutions coincide with the ESBH. It could be useful to check whether the action \eqref{eq:solutionofESBH} may arise from a different kind of reduction of 3D strings. 
\item 
Some applications require spacetime to be asymptotically $AdS$, so one may study a conformally transformed version of the ESBH which for large $\Phi$ approaches $AdS_2$, the ground state solution according to previous discussion. If the asymptotic value for curvature,
  \begin{equation}
    \label{eq:rgs}
    r_{\rm asy} = 2I^{-1}\frac{\extd}{\extd\Phi}\left(I^{-1}\frac{\extd}{\extd\Phi}\left(I\tilde{w}\right)\right)\,,
  \end{equation}
is required to be constant and the conformally invariant function $\tilde{w}$ in \eqref{eq:wI} behaves asymptotically like $\tilde{w}=-b\,\Phi^\beta$ with $\beta\neq 1$ then the non-invariant function must behave asymptotically as $I\propto\Phi^{\beta-2}$, which may be achieved by an appropriate $\Phi$-dependent conformal factor that is regular globally except for the asymptotic region $\Phi\to\infty$. More concretely, any two metrics $\tilde{g}_{\mu\nu}$ and $g_{\mu\nu}$ of the form \eqref{eq:EF} which differ only by the non-invariant function, $\tilde{I}$ and $I$, respectively, are connected by a conformal transformation $\tilde{g}_{\mu\nu}=\Om^2g_{\mu\nu}$ with $\Om^2=\tilde{I}/I$.
If the same procedure is applied for $\beta=1$ then \eqref{eq:rgs} not only leads to constant but to vanishing curvature. Unfortunately this applies to the ESBH as seen from \eqref{eq:wtilde} which for large $\Phi$ yields $\tilde{w}_+=-b\,\Phi + \dots$. Nevertheless, with the asymptotic choice
\begin{equation}
  \label{eq:confact}
  \left.\Om\right|_{\Phi\to\infty}\propto\frac{1}{\ln{\Phi}}
\end{equation}
for the conformal factor the ESBH may be transformed from the asymptotically flat frame \eqref{eq:line3}, \eqref{eq:killingnorm} to an asymptotically $AdS_2$ frame. The transformed version of \eqref{eq:integratingfactor} for large $\Phi$ behaves as $\tilde{I} \propto 1/(\Phi\ln^2{\Phi})+\dots$. The 
proportionality constant in \eqref{eq:confact} determines the asymptotic value of $r_{\rm asy}$. 
\elist

\subsection{Supersymmetrization, critical collapse and quantization}\label{se:matter}
 
Supersymmetrization \cite{Park:1993sd,Ikeda:1994dr,Strobl:1999zz} 
is possible if, and only if, the potential $V$ in \eqref{eq:pot} may be expressed in terms of a pre-potential $u$ such that
\begin{equation}
  \label{eq:prepotgen}
  V(\Phi)=-\frac18 \left(u^2(\Phi)U(\Phi)+\frac{\extd }{\extd \Phi}\, u^2(\Phi)\right)\,.
\end{equation} 
Once the pre-potential is defined one may apply a standard machinery \cite{Strobl:1999zz,Bergamin:2003mh} 
to obtain the supersymmetry transformations and all classical solutions including the BPS states. Because it is not the purpose of the present work to review these techniques the focus is solely on the pre-potential.
Provided the relation \eqref{eq:MGS} holds the pre-potential always exists.\footnote{This statement probably is not completely obvious although its derivation is simple: Inserting into the definitions \eqref{eq:wI}, \eqref{eq:prepotgen} and taking the positive root yields $u(\Phi)=\sqrt{-8\tilde{w}(\Phi)/I(\Phi)}$ (cf.~e.g.~(B.12) in ref.~\cite{Bergamin:2003mh}). For generic $\tilde{w}$ and $I$ a real pre-potential need not exist. However, if \eqref{eq:MGS} is true then the argument of the square root is non-negative. Note that under the shift $u^2(\Phi)\to u^2(\Phi)+\beta/I(\Phi)$ for arbitrary $\beta\in\mathbb{R}$ the potential $V$ is invariant. Thus, for a given bosonic model the pre-potential is not unique. This ambiguity corresponds to the freedom to choose $\tilde{w}^0$ in \eqref{eq:wtilde}. It has been fixed conveniently in \eqref{eq:prepot}. I am grateful to L.~Bergamin for correspondence on this subject.} For the ESBH ($+$)/ESNS ($-$) it reads
\begin{equation}
  \label{eq:prepot}
  u_\pm(\Phi_\pm) = -4\tilde{w}_\pm(\Phi_\pm) = 4b\left(1\pm\sqrt{\ga^2+1}\right)\,.
\end{equation}
It may be checked that the relations $\extd u^2/\extd \Phi=32b^2\ga$ and $u^2U=-16b^2\gamma$ are valid and thus the correct potential $V=-2b^2\ga$ is recovered from \eqref{eq:prepotgen}.
For sake of consistency the limiting cases may be studied at the level of the pre-potential. The weak coupling limit 
\begin{equation}
  \label{eq:weaku}
  \left.u_\pm\right|_{\ga\gg 1} \approx \pm 4b\,\ga \approx \pm 4b\,\Phi
\end{equation}
implies the correct pre-potential of the Witten BH, as expected. The strong coupling limit yields
\begin{equation}
  \label{eq:strongu}
  \left.u_+\right|_{\ga\ll 1} \approx 8b+2b\,\ga^2\approx 8b+b\,\Phi_+^2/2\,,\quad \left.u_-\right|_{\ga\ll 1} \approx -2b\,\ga^2 \approx -2b\, (6\Phi_-)^{2/3}\,. 
\end{equation}
In $u_+$ the next-to-leading order term has to be considered because otherwise a wrong result for $V$ is obtained since the leading order term is constant and therefore does not contribute to $\extd u^2_+/\extd \Phi_+$. Although the pre-potential $u_+$ is not the one of the JT model (which is linear in the dilaton), nevertheless to leading order $V$ as derived from $u_+$ displays the correct (linear) behavior. It is reassuring that indeed $u_-$ is the pre-potential for the spherically reduced 5D Schwarzschild BH.\footnote{For spherically reduced gravity from $D$ dimensions ($D>3$) the pre-potential reads $u(\Phi)\propto \Phi^{(D-3)/(D-2)}$. Inserting $D=5$ essentially yields $u_-$ in \eqref{eq:strongu}.} 

Coupling to matter degrees of freedom makes the theory non-topological in general. Integrability is lost apart from certain special cases (like the Witten BH or JT). Thus, dynamics is richer but also more complicated to describe. Already the addition to \eqref{eq:solutionofESBH} of the action for a single massless scalar field $\scalarf$,
\begin{equation}
  \label{eq:nonminscalar}
   S_{{\rm gen} KG}=\int_{\mathcal{M}_2} \extd^2x\sqrt{-g} \,F(\Phi)\, g^{\mu\nu}\left(\partial_\mu\scalarf\right)\left(\partial_\nu\scalarf\right)\,,
\end{equation}
which generalizes \eqref{eq:kgaction} slightly by allowing for nonminimal coupling to the dilaton via $F(\Phi)$, is capable to
introduce a new physical phenomenon: critical collapse. For instance, the spherically reduced Einstein-massless-Klein-Gordon model ($V=\rm const.$, $U=-1/(2\Phi)$, $F\propto\Phi$) leads to the famous Choptuik-scaling \cite{Choptuik:1993jv}
\begin{equation}
  \label{eq:choptuik}
  M_{BH} \propto (p-p_\ast)^\gamma\,,
\end{equation}
where $p\in[0,1]$ is a free parameter characterizing a one-parameter family of initial data with the property that for $p<p_\ast$ a BH never forms while for $p>p_\ast$ a BH always forms with mass $M_{BH}$ determined by \eqref{eq:choptuik} for $p$ sufficiently close to $p_\ast$. The critical parameter $p_\ast\in(0,1)$ may be found by elaborate numerical analysis and depends on the specific family under consideration; but the critical exponent $\gamma\approx 0.37$ is universal, albeit model dependent. Other systems may display a different critical behavior, cf.~the review ref.~\cite{Gundlach:1998wm}. The critical solution $p=p_\ast$ in general exhibits remarkable features, e.g.~discrete or continuous self-similarity and a naked singularity. 
It may be interesting to perform similar numerical studies for the ESBH coupled e.g.~to a scalar field \eqref{eq:kgaction}, \eqref{eq:sn1} or \eqref{eq:nonminscalar}, to obtain critical exponents and to study their dependence on the level $k$. 

A particular example of coupling to matter, namely to the tachyon, is now addressed with possible implications for 2D type 0A/0B string theory.
For the matrix model description of 2D type 0A/0B string theory cf.~\cite{Takayanagi:2003sm,Douglas:2003up} (for an extensive review on Liouville theory and its relation to matrix models and strings in 2D cf.~\cite{Nakayama:2004vk}; some earlier reviews are refs.~\cite{Ginsparg:1993is}; 
the matrix model for the 2D Euclidean string BH has been constructed in \cite{Kazakov:2000pm}). Although some of the subsequent considerations may have implications for matrix models their thorough discussion will not be attempted in the present work.
Recently \cite{Douglas:2003up,Thompson:2003fz} 
the low energy effective action for 2D type 0A/0B string theory in the presence of RR fluxes has been studied from various aspects. For sake of definiteness henceforth focus will be on 2D type 0A with an equal number $q$ of electric and magnetic D0 branes. The corresponding action in the second order formulation reads (remember \eqref{eq:dilatondilaton})
\begin{equation}
  \label{eq:sn20}
  S_{0\rm A}=\int_{{\mathcal M}_2} \extd^2x\sqrt{-g}\Big[X\frac{r}{2}-\frac{(\nabla{X})^2}{2X}+2b^2X - \frac{b^2q^2}{8\pi}\Big] + S_{\tac}\,,
\end{equation}
where $\tac$ denotes the tachyon.
The translation into the first order form is straightforward \cite{Grumiller:2004wi},
\begin{equation}
  \label{eq:sn21}
  I(X)=\frac{1}{X}\,,\quad \tilde{w}(X)= - 2 b^2 X  + \frac{b^2q^2}{8\pi}\ln{X}\,,
\end{equation}
because of
\begin{equation}
  \label{eq:potentialsRR}
  U(X)=-\frac1X \,,\quad V(X)=-2b^2X+\frac{b^2q^2}{8\pi} \,. 
\end{equation}
The action defining the tachyon sector up to second order in $\tac$ is given by
\begin{equation}
  \label{eq:sn1}
  S_{\tac} = \frac12 \int_{{\mathcal M}_2}\extd^2x\sqrt{-g}\left[ F(X) g^{\mu\nu} (\partial_\mu\tac)(\partial_\nu\tac) + f(X,\tac)\right]\,,
\end{equation}
with
\begin{equation}
  \label{eq:sn22}
  F(X)=X\,,\quad f(\tac,X) = b^2\tac^2\left(X-\frac{q^2}{2\pi}\right)\,.
\end{equation}
The geometric part in \eqref{eq:sn20} generalizes the Witten BH due to the inclusion of RR fluxes (compare \eqref{eq:potentialsRR} with \eqref{eq:potentialwittenbh}). The tachyon action \eqref{eq:sn1} introduces a PPDOF, the tachyon field. 
This suggests a twofold generalization of \eqref{eq:solutionofESBH}: one may simply add to $V$ in \eqref{eq:solutionofESBH3} a term corresponding to RR fluxes -- it might be just a constant proportional to $q^2$ as in the Witten BH limit \eqref{eq:potentialsRR} or a more complicated function depending on $\Phi$ which only for $k\to\infty$ approaches this constant. In principle the term could depend on $B$ as well, thus breaking dilaton-shift invariance discussed in \eqref{eq:dilshift}; in that case it is possible to circumvent the introduction of $q$ in the action by adjusting $\phi_0$ in \eqref{eq:solutionofESBH7} appropriately.\footnote{As demonstrated in appendix \ref{app:B.2} the parameter $q$ (or $b q$) may be eliminated from the action by ``integrating in'' a Maxwell field. Rather than introducing a new one the existing one may be exploited, e.g.~by adding to $V$ a term quadratic in $B$.} 
The second possibility involves no guess work and consists of the addition of the tachyon action \eqref{eq:sn1} to \eqref{eq:solutionofESBH} with $X$ replaced by $\ga B$, where $\ga$ can be expressed as a function of $\Phi$ by inverting \eqref{eq:solutionofESBH2.5}.  
This may be an interesting model on its own.\footnote{A remark is in order: according to the discussion in section 3 of \cite{Tseytlin:1993df} the one-loop form of the tachyon $\beta$-function is an exact result, provided metric and dilaton of the ESBH are the exact solutions derived from equating the corresponding $\beta$-functions to zero. Because the solutions of \eqref{eq:solutionofESBH} reproduce the ESBH (without involving non-linear or non-local field redefinitions), equation \eqref{eq:sn22} may be considered as an exact result for the effective Tachyon action, at least for $q=0$.} 
If $B$ is replaced by its on-shell value \eqref{eq:solutionofESBH7} it is evident that the constant $\phi_0$ now plays the role of a relative coupling constant between geometry and the tachyon sector. Therefore, as soon as the tachyon enters the game the abelian $BF$ term in \eqref{eq:solutionofESBH} ceases to be of purely auxiliar nature and acquires a physical status.

Finally, if one treats it as a model from scratch one may wish to quantize the ESBH action \eqref{eq:solutionofESBH} and/or its supersymmetrized version with pre-potential \eqref{eq:prepot}, possibly supplemented by matter degrees of freedom.
This was performed for generic supersymmetric models\footnote{Some ``bosonic references'' on path integral quantization of generic dilaton gravity with matter are \cite{Kummer:1998zs,Grumiller:2002nm}. 
In the absence of matter the theory is locally quantum trivial \cite{Kummer:1997hy}.} in ref.~\cite{Bergamin:2004us}, so in a sense the ESBH had been quantized before its action was constructed.

\acknowledgments{I would like to thank Dima Vassilevich for collaboration on the no-go result regarding an action for the ESBH, as well as for helpful discussion. I am grateful to Sergei Alexandrov for useful correspondence and for drawing Dima's and my attention to the ESBH 3 years ago. I thank Arkady Tseytlin for pointing out and discussing \cite{Tseytlin:1993df}. I am deeply indebted to Wolfgang Kummer for introducing me to the subject of 2D dilaton gravity 7 years ago. This work has been supported by an Erwin-Schr\"odinger fellowship, project J-2330-N08 of the Austrian Science Foundation (FWF). Some ideas presented in this work emerged during a visit at MIT in December 2004, and I am grateful to the CTP group for support and for an inspiring atmosphere, in particular to Mauro Brigante, Daniel Freedman, Alfredo Iorio, Roman Jackiw and Carlos Nu\~nez.
The final preparations of this paper have been performed just before the ``Wolfgangfest'' in January 2005 at the Vienna University of Technology, supported by project P-16030-N08 of the FWF.

\begin{appendix}

\section{No-go recap}\label{app:B.1}

The no-go result  obtained in \cite{Grumiller:2002md} relies upon the following assumption: the starting point has been an action \eqref{eq:FOG} with some generic function ${\mathcal V}$, i.e., not restricted to the simpler form \eqref{eq:pot}. Then it has been noted that the integration constant $\phi_0$ enters only the dilaton
field \eqref{eq:dvv3}, but not the metric \eqref{eq:dvv5}. Therefore, a symmetry property exists which
proved very important: a constant shift of the dilaton $\phi$ maps
a solution to another one of the same model. It has been shown next that this allows only two classes of ${\mathcal V}$:
\begin{equation}
  \label{eq:recap1}
  {\mathcal V}^{(1)} = V(X^+X^-)\,,\quad{\mathcal V}^{(2)} = X U \left(\frac{X^+X^-}{X^2}\right)
\end{equation}
For ${\mathcal V}^{(1)}$ dilaton shift invariance acted additively, i.e., $X\propto\phi$, while for ${\mathcal V}^{(2)}$ it acted multiplicatively, i.e., $X\propto \exp{(-2\al\phi)}$ with some non-vanishing $\al\in\mathbb{R}$. The first possibility has been excluded immediately,\footnote{While the conclusion is correct that no potential of type ${\mathcal V}^{(1)}$ reproduces the ESBH, the argument in \cite{Grumiller:2002md} is a bit too simple. But, along the lines of the no-go result for ${\mathcal V}^{(2)}$ one can provide a more elaborated argument that leads to the same conclusions for ${\mathcal V}^{(1)}$.} while the second one required further investigation. 
Actually, prospects for ${\mathcal V}^{(2)}$ did not seem bad at first glance, because both the Witten BH and the JT model are of that form (with $U(Z)=-Z-2b^2$ and $U(Z)=-b^2$, respectively). Nevertheless, by constructing the classical solutions and comparing with the ESBH it could be shown that for no value of $\alpha$ the whole family of ESBH solutions may be produced. 
An approximative model found in this way was interesting on its own and mimicked several important properties of the ESBH, but it was not ``the real stuff''. Still, several of the technical details spelled out in \cite{Grumiller:2002md} turned out to be very profitable for the construction of \eqref{eq:solutionofESBH}, in particular the ones contained in section 4.1 and appendix B of that work. Finally, it has been suggested in the outlook that the consideration of either non-localities or matter degrees of freedom could help to circumvent the no-go result. The introduction of the latter in general would imply additional PPDOF and thus a qualitative change as compared to the ``pure'' dilaton gravity case, where no PPDOF are present. 

There is an important exception to the ``rule'' that adding matter implies adding PPDOF: in two dimensions gauge fields do not carry PPDOF, which weakly suggests to add some gauge field(s) to \eqref{eq:FOG} in order to circumvent the no-go result without destroying the topological nature of the theory. However, there are infinitely many possibilities of adding gauge fields, so a supplementary selection criterion is needed. This is provided by the considerations below.

\section{The art of gauging constants}\label{app:B.2}

Suppose that $V$ in \eqref{eq:pot} depends on a real parameter $\lambda$, $V=V(X,\lambda)$. It is possible to eliminate it by ``integrating in'' an abelian $BF$ term such that on-shell $B=\lambda$ and the potential reads $V=V(X,B)$. If one encounters more parameters one can introduce a different abelian $BF$ term for each of them. This implies an additional abelian gauge symmetry for each parameter eliminated in this way. As a simple example the spherically reduced Schwarzschild BH may be considered. Its potentials read
\begin{equation}
  \label{eq:ssbh}
  U(X)^{(SSBH)}=-\frac{1}{2X}\,,\quad V(X)^{(SSBH)}=-\la^2\,.
\end{equation}
The scale parameter $\lambda$ is not very relevant because it just defines the physical units of the surface area. Still, one my opt to eliminate it from the action, which is possible by integrating in a $BF$ term. Thus, the spherically reduced Schwarzschild BH may be derived from a parameter free action of type \eqref{eq:fogu1} with\footnote{A brief digression: It seems that this additional $U(1)$ gauge symmetry arising in the Schwarzschild BH has not been addressed in detail before in the literature. While it is a somewhat trivial feature it may still be of use.}
\begin{equation}
  \label{eq:ssbh2}
  {\mathcal V}^{(SSBH)}(X^+X^-,X,B)=-\frac{X^+X^-}{2X}-B^2\,,
\end{equation}
where on-shell $B=\lambda$. 

This technique to the best of my knowledge was first introduced by Cangemi and Jackiw \cite{Cangemi:1992bj} while formulating the conformally transformed string inspired CGHS model as a gauge theory based upon the centrally extended Poincar{\'e} algebra,
\begin{equation}
  \label{eq:app1}
  [P_a,P_b]= \epsilon_{ab} \lambda I\,,\quad [P_a,J]=\epsilon_{ab}P^b\,,\quad [I,P_a]=[I,J]=0\,,
\end{equation}
where $P_a$ are generators of translation, $J$ generates boosts and $\lambda I:=Z$ is the central extension. Introducing the connection $\mathcal{A}=\om J+e^aP_a+A Z$ and Lagrange multipliers $X_A=(X,X_a,B)$ transforming under the coadjoint representation, the non-abelian $BF$ theory
\begin{equation}
  \label{eq:app2}
  S^{(CG)} = \int X_A \mathcal{F}^A\,,
\end{equation}
with $\mathcal{F}^A$ being the components of the curvature 2-form $\mathcal{F}=\extd\mathcal{A}+[\mathcal{A},\mathcal{A}]/2=\extd\om\, J+(\extd e^a+\eps^{ab}\om\wedge e_b)P_a+(\extd A-\epsilon)Z$, yields a particular Maxwell-dilaton gravity model \eqref{eq:fogu1} where\footnote{Compare e.g.~with (36) of ref.~\cite{Achucarro:1992mb}, noting that $\chi_A$ there corresponds to $X_A$ here. The construction of the sentence around this footnote -- a trifle too long for the hasty reader as it starts before \eqref{eq:app2} and ends after \eqref{eq:app5} -- is inspired by the vigor of Wolfgang's grammatical skills and a tribute to the good old days when sentences were allowed to frolic for a while before converging to a full stop.}
\begin{equation}
  \label{eq:app3}
  \mathcal{V}^{(CG)}=-B\,,
\end{equation}
a pre-cursor of which had been proposed earlier by Verlinde \cite{Verlinde:1991rf}, based upon the non-extended version of \eqref{eq:app1},
\begin{equation}
  \label{eq:app4}
  [P_a,P_b]=0\,,\quad [P_a,J]=\epsilon_{ab}P^b\,,
\end{equation}
leading to an action which in opposition to \eqref{eq:app2} explicitly depends on the parameter $\lambda$,
\begin{equation}
  \label{eq:app5}
  S^{(V)}=\int \left[X_A\mathcal{F}^A+2\lambda\epsilon\right]\,,
\end{equation}
where $X_A$ and $\mathcal{A}^A$ are the same as before except for the now absent Maxwell-component, $B$ and $A$, respectively. 

One can convince oneself immediately that integrating out $A$ in \eqref{eq:app2} yields $\extd B=0$, and setting $B=-2\lambda$ produces \eqref{eq:app5}. Thus, turning the argument around, one may ``integrate in'' a pair $A,B$ in order to eliminate a parameter from the potential ${\mathcal V}$. This generalizes readily to arbitrary theories of 2D dilaton gravity. Therefore, the appearance of a parameter in the action, like $\phi_0$ from \eqref{eq:dvv3}, need not present a problem because with the trick above it can be converted into a constant of motion at the cost of introducing an abelian $BF$ term. 

With this lesson in mind we reconsider the no-go result (see appendix \ref{app:B.1}) and the crucial ingredient to it, dilaton shift invariance (implemented multiplicatively). For the ESBH there are three parameters: $k$, $\phi_0$ and $b$. One (combination) of them must emerge as ``mass'', while the other ones may be either parameters of the action or, by integrating in $BF$ terms, constants of motion. We choose to keep $b$ as a parameter of the action and thus have to consider a single $BF$ term only. Starting with a Lagrangian of the form \eqref{eq:fogu1} one may require global invariance of the classical EOM under
\begin{equation}
  \label{eq:dvvnew2}
  X\to\la X\,,\quad X^a\to\la X^a\,,\quad B\to\la B\,,
\end{equation}
while the gauge fields $\om,e^a,A$ do not transform. This restricts to potentials of the type
\begin{equation}
  \label{eq:dvvnew3}
  \mathcal{V}^{(3)}=X U\left(\frac{X^aX_a}{B^2},\,\frac{X}{B}\right)\,,
\end{equation}
with some arbitrary two argument function $U$,
as opposed to the more constrained form of $\mathcal{V}^{(2)}$ in \eqref{eq:recap1}. But this means actually that $\mathcal{V}^{(3)}$ may be considered as arbitrary function of $X^aX_a$ and $X$ as long as appropriate factors of $B$ are attached. Thus, one of the restrictions that has been pivotal to the no-go result no longer is present. As it turned out, the lack of this restriction already was sufficient to construct the action \eqref{eq:solutionofESBH}. However, dilaton shift invariance is implemented in a slightly different way there:
\begin{equation}
  \label{eq:dilshift}
  X\to\la X\,,\quad X^a\to X^a\,,\quad B\to\la B\,,\quad A\to\la^{-1}A
\end{equation}
Therefore, the $X_aT^a$ term is invariant (rather than being multiplied with $\lambda$ as implied by \eqref{eq:dvvnew2}) and hence for consistency all other terms in the action should be invariant, too. This is possible if $X$ appears only in the combination $X/B$, which is indeed the case. Hence, dilaton shift invariance does not only leave the EOM invariant, but also the action \eqref{eq:solutionofESBH} and is therefore a global symmetry.

A final remark is in order: the action \eqref{eq:solutionofESBH} still depends on the parameter $b$, so by the same token as above one may introduce a second $BF$ term eliminating it. This second $BF$ term corresponds to the one introduced by Cangemi and Jackiw, while the one present in \eqref{eq:solutionofESBH} may be considered as eliminating $\phi_0$ from the action, as seen from the on-shell value of $B$ in \eqref{eq:solutionofESBH7}.

\end{appendix}

\providecommand{\href}[2]{#2}\begingroup\raggedright\endgroup


\end{document}